\documentclass[a4paper,11pt]{article}
\usepackage{jheppub} 
\usepackage{lineno}
\usepackage{braket}
\usepackage{amsmath}

\title{Exact formula for geometric quantum complexity of cosmological perturbations}

\author[a,b,1]{Satyaki Chowdhury,\note{Corresponding author.}}
\author[a]{Jakub Mielczarek,}

\affiliation[a]{Institute of Theoretical Physics, Jagiellonian University, Lojasiewicza 11, 30-348 Cracow, Poland}
\affiliation[b]{Doctoral School of Exact and Natural Sciences, Jagiellonian University, Lojasiewicza 11, 30-348 Cracow, Poland}

\abstract{Nielsen’s geometric approach offers a powerful framework for quantifying the complexity of unitary transformations. In this formulation, complexity is defined as the length of the minimal geodesic in a suitably constructed geometric space associated with the Lie group of relevant operators. Despite its conceptual appeal, determining geodesic distances on Lie group manifolds is generally challenging, and existing treatments often rely on perturbative expansions in the structure constants. In this work, we circumvent these limitations by employing a finite-dimensional matrix representation of the generators, which enables an exact computation of the geodesic distance and hence a precise determination of the complexity. We focus on the $\mathfrak{su}(1,1)$ Lie algebra, relevant for quantum scalar fields evolving on homogeneous and isotropic cosmological backgrounds. The resulting expression for the complexity is applied to de Sitter spacetime as well as to asymptotically static cosmological models undergoing contraction or expansion.}

\begin{document}
\maketitle
\flushbottom

\section{Introduction}
\label{sec:intro}

The study of quantum complexity emerged from quantum computation as a natural generalization 
of concepts from classical computational complexity. It serves as a mathematical framework 
for guiding the design of optimal algorithms for solving computational problems. At its core, 
quantum complexity addresses the fundamental question: \emph{How difficult is it to obtain a given quantum state?}

Over the past decade, complexity has become relevant for quantum gravity. Until then, entanglement has already established itself as a standard term in high-energy physics vocabulary due to the AdS/CFT correspondence, Ryu Takayanagi formula \cite{Ryu:2006bv} and its generalizations \cite{Hubeny:2007xt}. It had become extremely fruitful in elucidating the emergence of spacetime. However, black holes posed serious challenges in the interpretation of spacetime in terms of entanglement in the dual field theory. In particular, how the behavior of the interior of black holes, a region inaccessible to the Ryu-Takayanagi surfaces in asymptotically AdS spaces, is encoded in the dual CFT. In particular, the volume of the Einstein-Rosen bridge connecting the two sides of an eternal black hole keeps increasing through time exponentially in the black hole entropy, far after the timescale in the dual field theory where all entanglement-related quantities have equilibrated \cite{Susskind:2014moa}. This led to the search for quantities in the dual theory that capture this long-time growth of the Einstein-Rosen bridge on the gravity side. It was proposed that quantum complexity in the dual field theory is the quantity that could accurately capture the long-time growth of the Einstein-Rosen bridge in the dual field theory.  

Two conjectures connecting the complexity of the boundary field theory with geometrical quantities on the bulk side were proposed, the so-called ``complexity=volume" \cite{Stanford:2014jda} and ``complexity=action" \cite{Brown:2015bva}. The computation of holographic complexity on the bulk side was done in \cite{Carmi:2017jqz} and was extended in \cite{Swingle:2017zcd,Fu:2018kcp,Alishahiha:2018tep,Chapman:2016hwi,Yang:2017czx,Couch:2017yil}. Conceptually, quantum complexity is a measure of the minimal number of simple operations required to successfully implement a given transformation, say from a reference state to a desired target state. From the perspective of quantum circuits, one might think of it as the number of quantum gates present in the optimal circuit required to carry out the transformation. Nielsen et. al. geometerized the idea of finding optimal quantum circuits in a series of papers \cite{Nielsen_2006,https://doi.org/10.48550/arxiv.quant-ph/0502070,https://doi.org/10.48550/arxiv.quant-ph/0701004}. They established a correlation between quantum complexity and the length of minimal geodesics in the space of unitary operations. This geometric definition enabled the use of differential geometry ideas in the study of quantum complexity. Nielsen's original formulation was based on determining the optimal circuits for $n$-qubit operations. An accurate verification of the correspondence required a proper definition of complexity in quantum field theory. Several notable efforts were made to generalize the idea put forward by Nielsen et. al to quantify the complexity of individual states in quantum mechanics and quantum field theories \cite{Jefferson:2017sdb,Khan:2018rzm,Bhattacharyya:2018bbv,Chapman:2018hou,Caceres:2019pgf,Hackl:2018ptj}. It was first put forward for free scalar fields in \cite{Jefferson:2017sdb} and was extended for fermionic fields\cite{Khan:2018rzm} and interacting scalar fields in \cite{Bhattacharyya:2018bbv}, where the authors defined and calculated the complexity of ground state in the field theory. These works were extended in several different contexts in \cite{Chapman:2017rqy,Chapman:2018hou,Caceres:2019pgf,Camargo:2018eof,Ali:2018aon,Ali:2019zcj}. The basic idea behind these works was to choose a reference state and a set of quantum operators or gates. The complexity of the target state is then the length of the minimal circuit connecting the reference state to the target state by successive applications of the chosen gates (quantum operators). In practice, most of these attempts were restricted to the requirement of Gaussian reference and target states. However, such restrictions are not ideal in field theories and interacting systems. Another limitation of ``state complexity" is the requirement of defining an arbitrary reference state. The choice of quantum operators or ``gates" is then determined by the choice of reference state. This makes the complexity of the target state arbitrarily dependent on the choice of the reference state and hence arbitrary. 

Quantifying the complexity of individual quantum states has its own significance in the context of quantum simulations in many-body quantum mechanics and in quantum gravity. The complexity of the thermofield double state is one such example, which is dual to an eternal black hole in AdS. The complexity of the thermofield double state was studied in \cite{Chapman:2017rqy}. Although the complexity of individual states represents remarkable progress, it is certainly not the desirable one to provide a comprehensive understanding of the complexity of quantum processes in general. A more suited approach for quantifying the complexity of dynamical quantum processes is the notion of ``operator complexity" that allows us to deal with the time evolution operator directly without the need to specify any initial reference Gaussian state.

Although, the ``state'' and ``unitary (operator)'' complexities are two different notions, they can be related. State complexity can also be interpreted as the complexity of the least complex unitary connecting the reference and the target states. The connection between these two different approaches to complexity can be written as:
\begin{align}
    C\bigg[\ket{\Psi}_R\rightarrow \ket{\Psi}_T\bigg]= \text{min}_{\hat{U}}C\left[\hat{U}\right],
\end{align}
where $\hat{U}$ is a unitary transformation that takes $\ket{\Psi}_R$ to $\ket{\Psi}_T$. 
While, in operator complexity, the desired unitary is the standalone object, in the state 
complexity approach there can multiple unitary transformations that connects the reference and 
the target states. This number depends on the choice of the reference state and the quantum gates. 
The advantage of operator complexity over state complexity is that the requirement of defining 
an arbitrary reference state is removed. For operators, by default, it is always the identity operator.
Of course, one can consider something like the relative complexity between two unitary 
operators $\hat{U}_1$ and $\hat{U}_2$, but this can be translated to the complexity of 
$\hat{U}_1\hat{U}_2^{-1}$. Also, in operator complexity, the choice of fundamental operators 
(quantum gates) is directly related to the target unitary operator.

An extension of Nielsen's general methodology to investigate the complexity of dynamical quantum systems requires replacing the unitary group $SU(2^N)$ (suitable for $N$ qubit systems) with other Lie groups suited for the purpose. For example, for harmonic oscillator systems, the groups $SU(1,1)$ and $Sp(2,R)$ play a prominent role. The methodology developed by Nielsen for $N$ qubit systems relied on finding geodesics on the $SU(2^N)$ group manifold. It was recently exploited to study complexity growth in the SYK model \cite{Balasubramanian:2018hsu, Balasubramanian:2021mxo}; a typical quantum chaotic system. In the geometrical formulation, the quantum complexity of a unitary operator is the length of the minimal geodesic on the unitary group manifold connecting identity to the unitary operator. One begins by identifying a set of fundamental operators ($\hat{\mathcal{O}}_I$) that are related to the unitary operator in some way. In other words, these fundamental operators will act as the generators of the quantum gates that will construct the circuit to realize the desired $\hat{U}$. Particularly, the Lie algebra generated by these fundamental operators can be exponentiated to a group of which the target unitary is an element. The identified fundamental operators are then classified as ``easy" or ``hard" by introducing the so-called \textit{penalty factor matrix} ($G_{IJ}$), which acts as the metric on the group manifold. This right-invariant metric equipped with the appropriate cost factors accurately captures the hardness of an operator by penalizing the motion along the ``hard" directions. Typically, the operators involving lower-order terms are assigned lower penalties compared to the ones involving higher-order terms \cite{Chowdhury:2023iwg}. This notion is particularly relevant when dealing with anharmonic or interacting systems \cite{Chowdhury:2023iwg,Bhattacharyya:2024rzz}. A choice of $G_{IJ} \neq \delta_{IJ}$ assumes different penalties or operational costs in different directions. This introduces anisotropy in the operator space geometry. The geodesics on Lie groups equipped with a right invariant metric can be found by solving the so-called \textit{Euler-Arnold} equation, which is given by \cite{AIF_1966__16_1_319_0,Balasubramanian:2019wgd}:
\begin{align}
\label{eulerarnold}
    G_{IJ}\frac{dV^J(s)}{ds}= f_{IJ}^K V^J(s) G_{KL}V^L(s),
\end{align}
where $f_{IJ}^{K}$ are the structure constants of the Lie algebra, defined by:
\begin{align}
    [\hat{\mathcal{O}}_I,\hat{\mathcal{O}}_J]= i f_{IJ}^K \hat{\mathcal{O}}_K.
\end{align}

The components $V^I(s)$ represent the tangent vector at each point along the geodesic, defined by:
\begin{align}
\label{differentialU}
    \frac{d\hat{U}(s)}{ds}= -i V^I(s)\hat{\mathcal{O}}_{I}\hat{U}(s).
\end{align}
Given a solution $V^I(s)$, an integration of \ref{differentialU}, results in the trajectory in the group manifold, guided by the velocity vector $V^I(s)$. This solution can be written as:
\begin{align}
\label{pathorderedexp}
    \hat{U}(s)= \mathcal{P}\exp\bigg(-i \int_{0}^s ds' V^I(s') \hat{\mathcal{O}}_I\bigg).
\end{align}
The boundary conditions:
\begin{align}
    \hat{U}(s=0)= \hat{\mathbb{I}}, ~~~~ \hat{U}(s=1)= \hat{U}_{\rm target},
\end{align}
are then imposed in order to filter out the trajectories that start from $\hat{\mathbb{I}}$ 
and reach $\hat{U}_{\rm target}$. Generally, \ref{eulerarnold}, defines a family of geodesics 
on the unitary group manifold. The boundary condition at $s=1$ filters out those geodesics that 
realizes the target unitary operator by fixing the magnitude of the tangent vector $V^I(s)$ at $s=0$. 
In other words, the boundary condition at $s=1$, fixes the initial velocity required to reach 
$\hat{U}_{\rm target}$. In principle, there might be more than one value of the initial velocities 
for which $\hat{U}_{\rm target}$ is reached. The optimal circuit realizing $\hat{U}_{\rm target}$ 
is then given by the shortest geodesic, and its length is defined as the complexity:
\begin{equation}
\label{eqn:complexityexpression}
	C[\hat{U}_{\rm target}] := {\rm min}_{\{V^I(s)\}}\int_{0}^{1}ds\sqrt{G_{IJ}V^{I}(s)V^{J}(s)},
\end{equation}
where the minimization is over all geodesics $\{V^I(s)\}$ from identity to $\hat{U}_{\rm target}$.

The motivations of the paper are as follows:
\begin{itemize}
    \item Firstly, we aim to generalize the upper bound result of the complexity of cosmological perturbations provided in \cite{Chowdhury:2024ntx}. The derived results were based on truncating the path-ordered exponential written in \ref{pathorderedexp} to the leading order term in the Dyson series, resulting in the interpretation of our results as \textit{upper bounds} on complexity instead of the actual value. Although the upper bound result provided insightful predictions, the reliability of those predictions ultimately depended on understanding how closely the upper bound approximates the true complexity. The time evolution operator relevant to cosmological perturbations can be constructed from the elements of the $\mathfrak{su}(1,1)$ Lie algebra. Therefore, the geometric complexity for the evolution of these perturbations is based on the finding of geodesics in the unitary group manifold formed by the generators of $\mathfrak{su}(1,1)$. In this paper, we used a well-known finite-dimensional matrix representation of the generators of $\mathfrak{su}(1,1)$ to derive the complexity of the time evolution operator. The use of a finite-dimensional matrix representation allows us to avoid dealing with the path-ordered exponential and work directly with the differential equation it satisfies. Considering the finite-dimensional representation of generators of well-known Lie algebras is a very popular technique in mathematical physics and quantum optics, and is frequently used in disentangling exponential operators and so on \cite{Mufti, Gilmore1, Ban1,Ban:93,gilmore2006lie}. 

    \item To verify the validity of the complexity formula derived from the matrix representation of the generators of $\mathfrak{su}(1,1)$, we prove that it satisfies the triangle inequality. Finally, we revisit the problem of the complexity of the time evolution operator of a scalar field mode on deSitter background and do a comparative analysis with the upper bound result derived in \cite{Chowdhury:2024ntx}.

    \item Furthermore, we study a model of a scalar field in an asymptotically static universe \cite{Birrell:1982ix}. This refers to a universe that is characterized by constant scale factors in the asymptotic past and future. Two situations are possible, an expanding universe, when the universe expands from a small initial scale factor $a_i$ to a large value $a_f$ and the contracting universe, where the opposite happens.

\end{itemize}

The organization of this article is as follows: The basic formalism of quantum fields on FRW spacetime is reviewed in Sec.~\ref{sec:sec2}. The structure of the time evolution operator, which is the product of the two-mode squeezed and rotation operators, is established in this section. In Sec.~\ref{sec3}, we first argue that the generators suitable to construct the desired target unitary operators satisfy the $\mathfrak{su}(1,1)$ Lie algebra. Using a finite-dimensional matrix representation of the $\mathfrak{su}(1,1)$ generators, we derive the formula for the geometric complexity. The formula for the relative complexity of between two such operators characterized by different squeezing parameters was derived in Sec.~\ref{sec4}. A proof of the formula satisfying the triangle inequality was also established in this section. In Sec.~\ref{sec5}, the complexity of the time evolution of a scalar field mode in de Sitter spacetime is discussed, followed by the investigation of complexity in an asymptotic static universe in Sec.~\ref{sec6}. 
Section~\ref{sec7} summarizes the key results and conclusions of the paper. 

\section{Quantum scalar field in FRW spacetime}
\label{sec:sec2}

In this section, we will briefly review the quantum scalar field theory on 
cosmological background. 

Let us consider the action of a minimally coupled massive scalar field $\phi$ on the flat FRW universe: 
\begin{align}
    S_{\phi} = -\int d^4 x \sqrt{-g}\bigg(\frac{1}{2}g^{\mu \nu} \partial_{\mu}\phi\partial_{\nu}\phi+ \frac{1}{2}m^2\phi^2\bigg),
\end{align}
with the metric given by: 
\begin{align}
     ds^2= g_{\mu\nu}dx^{\mu}dx^{\nu} = a^2(\tau)(-d\tau^2+d{\bf x}^2),
\end{align}
where $\tau$ is the so-called \emph{conformal time}, used for further convenience, 
and $a(\tau)$ is a scale factor.  

Introducing an auxiliary field $\chi= a\phi$, the action can be rewritten in the Minkowski-like form: 
 \begin{align}
  \label{eqn:quenchproof}
      S_{\chi} = \int d\tau L_{\chi}  = - \frac{1}{2}\int d\tau d^3 x \bigg(\frac{1}{2}\eta^{\mu \nu} \partial_{\mu}\chi \partial_{\nu}\chi+ M^2(\tau) \chi^2 \bigg),
 \end{align}
where $L_{\chi}$ is a Lagrange function and $\eta_{\mu\nu}$ is the Minkowski metric and the 
effective mass:
\begin{align}
    M^2(\tau) := m^2 a(\tau)^2 -\frac{a''}{a}.
\end{align}

At this stage, it is useful to pass to the 
Hamiltonian formalism. For this purpose, the 
momentum conjugated to $\chi$ is introduced:
\begin{equation}
\Pi : = \frac{\delta L_{\chi}}{\delta \chi}= \chi', 
\end{equation}
so that the following Poisson bracket holds:
\begin{equation}
\{ \chi({\bf x},\tau), \Pi({\bf y},\tau) \} = \delta^{(3)}({\bf x-y}).
\label{Poisson1}
\end{equation}

This allows us to perform the Legendre transform, which leads to the Hamilton function:
\begin{align}
    H_{\chi} = \frac{1}{2}\int d^3x 
    \bigg(\Pi^2 + (\nabla \chi)^2+ M^2(\tau) \chi^2\bigg). 
\end{align}

In what follows, we adopt the Heisenberg picture to quantize the scalar field. 
In this formulation, all time dependence is carried by the operators, while the 
states remain fixed. This choice is particularly well-suited to our analysis, 
as our focus is on the complexity of operators.

By promoting the $\chi$ and $\Pi$ to the 
quantum operators, the Poisson bracket 
(\ref{Poisson1}) leads to the following 
commutation relation: 
\begin{equation}
\left[ \hat{\chi}({\bf x},\tau), \hat{\Pi}({\bf y},\tau) \right] =  i \hat{\mathbb{I}}\delta^{(3)}({\bf x-y}).
\label{Bracktet1}
\end{equation}

The Hamiltonian operator can then be written as:
\begin{align}
    \hat{H}_{\chi}= \frac{1}{2}\int d^3x \bigg(\hat{\Pi}^2 + (\nabla \hat{\chi})^2+ M^2(\tau) \hat{\chi}^2\bigg). 
\end{align}
It will be useful to express the Hamiltonian in the 
Fourier space, for which we perform a Fourier transform 
of the field variable and the conjugate momentum:
\begin{align}
\label{eqn:fieldfourier}
    \hat{\chi}({\bf x},\tau) &= \int \frac{d^3 k}{(2\pi)^{3/2}} \hat{\chi}_{\bf k}(\tau) e^{i{\bf k} \cdot {\bf x}}, \\
     \hat{\Pi}({\bf x},\tau) &= \int \frac{d^3k}{(2\pi)^{3/2}} \hat{\Pi}_{\bf k}(\tau) e^{i{\bf k} \cdot {\bf x}}.
\end{align}

The Hamiltonian operator in the Fourier space can then be 
expressed as:
\begin{align}
\label{Hamiltonian}
    \hat{H}(\tau)= \frac{1}{4}\int d^3 k
\bigg[(\hat{\Pi}_{{\bf k}}\hat{\Pi}^{\dagger}_{{\bf k}}+\hat{\Pi}^{\dagger}_{\bf k}\hat{\Pi}_{{\bf k}})+ \omega_k^2(\tau) (\hat{\chi}_{{\bf k}}\hat{\chi}^{\dagger}_{{\bf k}}+\hat{\chi}^{\dagger}_{{\bf k}}\hat{\chi}_{{\bf k}})\bigg],
\end{align}
where $\omega_k^2(\tau) :=k^2+M^2(\tau)$, and $k^2 = {\bf k} \cdot {\bf k}$.

Time evolution of the Fourier components $\hat{\chi}_{\bf k}(\tau)$ and 
$\hat{\Pi}_{\bf k}(\tau)$ can be parametrized with the use of the so-called m
ode functions $f_k(\tau)$ and $g_k(\tau)$, so that:
\begin{align}
    \hat{\chi}_{\bf k} (\tau) &= f_k(\tau)\hat{c}_{\bf k}+ f^*_k(\tau)\hat{c}_{-{\bf k}}^{\dagger}, \\
    \hat{\Pi}_{\bf k}(\tau) &= g_k(\tau)\hat{c}_{\bf k}+ g^*_k(\tau)\hat{c}_{-{\bf k}}^{\dagger},
\end{align}
where $\hat{c}_{\bf k}$ and $\hat{c}_{\bf k}^{\dagger}$ 
are annihilation and creation operators at some initial 
time $\tau_0$, satisfying the standard commutation relation:
\begin{equation}
\left[ \hat{c}_{\bf k}, \hat{c}_{\bf q}^{\dagger} \right] = 
\hat{\mathbb{I}}\delta^{(3)}({\bf k-q}).
\label{BracktetC}
\end{equation}

The annihilation operator allows us to introduce the vacuum state 
$\ket{0}$, defined such that $\hat{c}_{\bf k} \ket{0}=0$. The 
Hilbert space of the system can then be constructed by 
subsequent actions of the creation operator $\hat{c}^{\dagger}_{\bf k}$
on the vacuum state, with all possible values of the vector ${\bf k}$.
The obtained basis states are in the form:
\begin{equation}
\ket{n_1,n_2,...}  = \prod_{i\in \mathbb{N}} \frac{\left( \hat{c}^{\dagger}_{{\bf k}_{i}} \right)^{n_i}}{\sqrt{n_i!}}       \ket{0},
\end{equation}
so that the Hilbert space is:
\begin{equation}
\mathcal{H} = \text{span}\left( \ket{n_1,n_2,...}  \right).
\end{equation}

Furthermore, conservation of the bracket (\ref{Bracktet1})
implies the so-called Wronskian condition:
\begin{equation}
f_kg_k^*-f^*_kg_k=i.
\end{equation}

Employing Hamilton's equations for the 
Fourier component operators, we find:
\begin{align}
    \frac{d\hat{\chi}_{\bf k}}{d\tau} &=-i [\hat{\chi}_{\bf k},\hat{H}] = \hat{\Pi}_{\bf_k}, \\
    \frac{d \hat{\Pi}_{\bf_k}}{d\tau}&=-i [\hat{\Pi}_{\bf_k},\hat{H}] = - \omega^2_k \hat{\chi}_{\bf k},
\end{align}
which implies that: 
\begin{align}
 g_k &= f'_k, \\
 \frac{d^2f_k}{d\tau^2}+\omega^2_k(\tau)f_k&=0,
\end{align}
where the second equation is the so-called equation 
of modes. 

The time evolution of the creation and annihilation 
operators can be expressed as the so-called Bogoliubov 
transformation:
\begin{align}
    \hat{c}_{\bf k}(\tau) &= \alpha_k^*(\tau)\hat{c}_{\bf k}- \beta_k^* (\tau) \hat{c}_{-\bf k}^{\dagger}, \\
    \hat{c}_{\bf k}^{\dagger}(\tau) &= \alpha_k(\tau)\hat{c}_{\bf k}^{\dagger}- \beta_k (\tau) \hat{c}_{-\bf k},
\end{align}
where the Bogoliubov coefficients $\alpha_k, \beta_k
\in \mathbb{C}$, satisfy the following normalization condition:
\begin{align}
    |\alpha_k|^2-|\beta_k|^2=1.
\end{align}

This ensures that the commutation relation between the creation 
and the annihilation operator is preserved in time. If we choose the initial state to be the vacuum at $\tau=\tau_0$, it implies 
that the Bogoliubov coefficients at $\tau_0$ are fixed to be:
\begin{align}
    \alpha_k(\tau_0)= e^{i\theta_0}, ~~~~~ \beta_k(\tau_0)= 0,
\end{align}
with some unspecified angle $\theta_0$.
 
The normalization condition satisfied by the Bogoliubov coefficients 
allows us to parameterize it hyperbolically as follows:
\begin{align}
\label{alphaandbeta}
    \alpha_k(\tau)= e^{-i\theta_k(\tau)}\cosh(r_k(\tau)), ~~~~ \beta_k(\tau)= e^{-i(\phi_k(\tau)-\theta_k(\tau))}\sinh(r_k(\tau)), 
\end{align}
where $r_k$, $\theta_k$, and $\phi_k$ are time-dependent parameters and are 
called the squeezing amplitude, rotation angle, and squeezing angle, respectively \cite{Schumaker:1986tlu,Caves:1985zz}.

Using the hyperbolic parametrization of the Bogoliubov coefficients, the creation and the annihilation operators at any time $\tau$ can be written as:
\begin{align}
    \hat{c}_{\bf k}(\tau) &= e^{i\theta_k(\tau)}\cosh(r_k(\tau))\hat{c}_{\bf k}- e^{i(\phi_k(\tau)-\theta_k(\tau))}\sinh(r_k(\tau))\hat{c}^{\dagger}_{-\bf k} \\
    \hat{c}_{\bf k}^{\dagger}(\tau) &= e^{-i\theta_k(\tau)}\cosh(r_k(\tau))\hat{c}_{\bf k}^{\dagger}- e^{-i(\phi_k(\tau)-\theta_k(\tau))}\sinh(r_k(\tau))\hat{c}_{-\bf k},
\end{align}

The above equations can be recast in the following form:
\begin{align}
    \hat{c}_{\bf k}(\tau) &= \hat{R}_2^{\dagger}(\theta_k)\hat{S}_2^{\dagger}(r_k,\phi_k)~\hat{c}_{\bf k}~ \hat{S}_2(r_k,\phi_k)\hat{R}_2(\theta_k),\\
    \hat{c}_{\bf k}^{\dagger}(\tau) &= \hat{R}_2^{\dagger}(\theta_k)\hat{S}_2^{\dagger}(r_k,\phi_k)~\hat{c}^{\dagger}_{\bf k}~ \hat{S}_2(r_k,\phi_k)\hat{R}_2(\theta_k),
\end{align}
where $\hat{R}_2(\theta)$ and $\hat{S}_2(r,\phi)$ are known as the two-mode 
rotation and squeezing operators \cite{Albrecht:1992kf} and are defined as:
\begin{align}
    \hat{R}_2(\theta) &:= \exp\bigg\{i \theta_k(\hat{c}^{\dagger}_{\bf k}\hat{c}_{\bf k}+ \hat{c}^{\dagger}_{-{\bf k}}\hat{c}_{-{\bf k}}+1)\bigg\}, \\
    \hat{S}_2(r,\phi) &:= \exp\bigg\{r_k (e^{-i\phi}\hat{c}_{\bf k}\hat{c}_{-{\bf k}}-e^{i\phi}\hat{c}^{\dagger}_{\bf k}\hat{c}^{\dagger}_{-{\bf k}})\bigg\}.
\end{align}
Please notice that the usual $1/2$ factor used in the definition of the squeezing parameter has been absorbed into the squeezing amplitude $r_k$.

The squeezing and the rotation operators are extensively used in the context 
of quantum optics \cite{Schumaker:1986tlu, Caves:1985zz}. It has also been 
significantly applied in the context of cosmology in \cite{Albrecht:1992kf, 
Grishchuk:1990bj, Grishchuk:1989ss,Grishchuk:1994sj,Martin:2004um,Martin:2007bw,
Lemoine:2008zz}

The above equations imply that the time evolution operator is a product of the two-mode squeezing and rotation operator. We will denote the time evolution operator as $U_{\rm evolution}$ from here on:
\begin{align}
    \hat{U}_{\rm evolution}= \hat{S}_2(r_k,\phi_k)\hat{R}_2(\theta_k).
\end{align}

By comparing the field operators' decompositions at $\tau_0$ and some 
arbitrary time $\tau$ we find:
\begin{align}
\label{modefunctions}
    f_k(\tau_0) &= \alpha_k^* (\tau) f_k(\tau)-\beta_k(\tau)f_k^*(\tau), \\\
    g_k(\tau_0) &= \alpha_k^*(\tau)g_k(\tau)-\beta_k(\tau)g_k^*(\tau).
\end{align}

Solving \ref{modefunctions} and using the Wronskian condition, the Bogoliubov coefficients can be expressed as functions of the mode functions, and its derivatives as:
\begin{align}
\label{Bogoliubovcoefficient}
    \alpha_k(\tau) &= -i (f_k(\tau)g^*(\tau_0)-f_k^*(\tau_0)g_k(\tau)), \\
    \beta_k(\tau) &= i(f_k(\tau)g(\tau_0)-f_k(\tau_0)g_k(\tau)).
\end{align}

The above equations allow us to express the squeezed parameter ($r_k$), the squeezing angle ($\phi_k)$ and the rotation angle ($\theta_k$) with the mode functions through Eq. \ref{alphaandbeta}:
\begin{align}
    r_k(\tau) &= {\rm arcsinh}|\beta_k(\tau)|,\\
    \theta_k(\tau) &= -{\rm arg} (\alpha_k(\tau)), \\
    \phi_k(\tau) &= {\rm arg}(\alpha_k(\tau)\beta_k(\tau)).
\end{align}

It is important to note that both the two-mode squeezing operator 
$\hat{S}_2$ and the two-mode rotation operator $\hat{R}_2$ can be 
generated with the use of the operators: 
\begin{align}
    \hat{\mathcal{O}}_1 &= \frac{1}{2}\left(\hat{c}_{\bf k}\hat{c}_{-{\bf k}}+ \hat{c}^{\dagger}_{{\bf k}}\hat{c}^{\dagger}_{-{\bf k}}\right), \\
    \hat{\mathcal{O}}_2 &= \frac{i}{2}\left(\hat{c}_{{\bf k}}\hat{c}_{-{\bf k}}- \hat{c}^{\dagger}_{\bf k}\hat{c}^{\dagger}_{-{\bf k}}\right), \\
    \hat{\mathcal{O}}_3 &= \frac{1}{2}\left(\hat{c}_{\bf k}^{\dagger}\hat{c}_{\bf k}+ \hat{c}^{\dagger}_{-{\bf k}}\hat{c}_{-{\bf k}}+1\right),
\end{align}
which satisfy the following $\mathfrak{su}(1,1)$ Lie algebra:
\begin{align}
    [\hat{\mathcal{O}}_1,\hat{\mathcal{O}}_2]= -i \hat{\mathcal{O}}_3 , ~~~~ [\hat{\mathcal{O}}_1,\hat{\mathcal{O}}_3]= -i \hat{\mathcal{O}}_2, ~~~~ [\hat{\mathcal{O}}_2,\hat{\mathcal{O}}_3]= i \hat{\mathcal{O}}_1.
\end{align}
So that:
\begin{align}
\hat{R}_2(\theta) &= \exp (2i\theta_k \hat{\mathcal{O}}_3), \\\
S_2(r,\phi) &= \exp\bigg[r_k(e^{-i\phi_k}(\mathcal{O}_1-i\mathcal{O}_2)- e^{i\phi_k}(\mathcal{O}_1+i\mathcal{O}_2))], \\
    &= \exp\bigg[-2ir_k (\sin(\phi_k)\mathcal{O}_1+ \cos(\phi_k)\mathcal{O}_2)\bigg].
\end{align}
This observation turns out to be crucial for the analysis 
of the quantum complexity in Nielsen's approach. 

\section{Geometric complexity in expanding spacetimes}

In this section, we study the complexity of the following objects:
\begin{itemize}
    \item The complexity of the \emph{out} vacuum relative to the \emph{in} vacuum.
    \item The complexity of the time-evolution operator.
\end{itemize}

To analyze these quantities, we examine the complexity of three unitary 
operators: the squeezing operator, the rotation operator, and their product. 
The latter corresponds to the full time-evolution operator.

\subsection{Complexity of a unitary operator belonging to $\mathfrak{su}(1,1)$}

The Lie group SU(1,1) is defined as the set of $2 \times 2$ matrices $U$ of determinant 1 satisfying the relation:
\begin{align}
    U \epsilon U^{\dagger}= \epsilon,
\end{align}
where:
\begin{align}
    \epsilon= \begin{pmatrix}
        1 && 0 \\
        0 && -1
    \end{pmatrix}.
\end{align}

With this matrix representation, a general group element can be written as: 
\begin{align}
    U(s)= \begin{pmatrix}
        \alpha(s) && \beta(s) \\
        \Bar{\beta}(s) && \Bar{\alpha}(s)
    \end{pmatrix},
    ~~~~~ {\rm with} ~~~~~ |\alpha|^2-|\beta|^2=1.
\end{align}

The Lorentzian Pauli matrices provide a widely used $2 \times 2$ 
matrix representation of the generators of the $\mathfrak{su}(1,1)$ 
algebra. We will adopt this representation for our purposes as well 
\cite{Livine:2012mh}:
\begin{align}
\label{matrixgenerators}
\mathcal{O}_1= \frac{1}{2}
\begin{pmatrix}
    0 && 1 \\
    -1 && 0
\end{pmatrix}
, ~~ \mathcal{O}_2= -\frac{i}{2}
\begin{pmatrix}
    0 && 1 \\
    1 && 0
\end{pmatrix}
, ~~ \mathcal{O}_3= \frac{1}{2}
\begin{pmatrix}
    1 && 0 \\
    0 && -1
\end{pmatrix}.
\end{align}

The advantage of using a finite-dimensional matrix representation of the generators is that it simplifies finding a solution to:
\begin{align}
    \frac{dU}{ds}=-i V^I(s)\mathcal{O}_IU(s). 
    \label{differentialU2}
\end{align}

However, one must keep in mind that the existence of a finite-dimensional matrix of the generators is not always guaranteed, particularly when dealing with interacting or anharmonic systems. 

Substituting the general element $U(s)$ and the $\mathcal{O}_I$'s 
in \ref{differentialU2}, we arrive at the following differential 
equations for $\alpha$ and $\beta$:
\begin{align}
    \alpha'(s) &= -\frac{i}{2}v_3 \alpha(s) - \frac{i}{2}(v_1-i v_2)(\cos(2 s v_3)-i \sin(2 s v_3)) \Bar{\beta}(s) \\
    \Bar{\beta}'(s) &= \frac{i}{2}(v_1+i v_2)(\cos(2 s v_3)+i \sin(2 s v_3)) \alpha(s)+ \frac{i}{2}v_3 \Bar{\beta}(s). 
\end{align}
Furthermore, a solution to the Euler-Arnold equation discussed in the Appendix \ref{appendixupperbound} has been used. 

We can reparametrize $\alpha$ and $\beta$ such that:
\begin{align}
    \alpha(s)= \cosh(\rho(s))e^{i\chi(s)}, ~~~ \beta(s)= -\sinh(\rho(s))e^{i\psi(s)},
\end{align}
so that the $U(s)$ matrix can be written as:
\begin{align}
\label{genericelement}
    U(s)= \begin{pmatrix}
        \cosh(\rho(s))e^{i\chi(s)} && -\sinh(\rho(s))e^{i\psi(s)} \\
        -\sinh(\rho(s))e^{-i\psi(s)} && \cosh(\rho(s))e^{-i\chi(s)}
    \end{pmatrix}.
\end{align}
with this parametrization, the differential equations for $\alpha$ and $\beta$ 
can be recast in terms of the equations for $\rho$, $\chi$ and $\psi$, which 
are as follows:
\begin{align}
    \rho'(s) &=\frac{1}{2} (v_1 \sin (2 s v_3+\chi (s)+\psi (s))+v_2 \cos (2 s v_3+\chi (s)+\psi (s))), \\
    \chi'(s) &=\frac{1}{2} (\tanh (\rho(s)) (v_1 \cos (2 s v_3+\chi (s)+\psi (s))-v_2 \sin (2 s v_3+\chi (s)+\psi (s)))-v_3), \\
    \psi '(s) &=\frac{1}{2} (\coth (\rho(s)) (v_1 \cos (2 s v_3+\chi (s)+\psi (s))-v_2 \sin (2 s v_3+\chi (s)+\psi (s)))-v_3).
\end{align}

By introducing:
\begin{align}
\label{ydefinition}
    y(s) := 2 s v_3+\chi(s)+\psi(s)
\end{align}
we can re-express the above equations as:
\begin{align}
    \rho'(s) &= \frac{1}{2} \bigg(v_1 \sin (y)+v_2 \cos (y)\bigg), \\
    \label{eqnchi}
    \chi'(s) &= \frac{1}{2} \bigg(\tanh (\rho(s)) (v_1 \cos (y)-v_2 \sin (y))-v_3\bigg), \\
    \label{eqnpsi}
    \psi'(s) &= \frac{1}{2} \bigg(\coth (\rho(s)) (v_1 \cos (y)-v_2 \sin (y))-v_3\bigg).
\end{align}

Now, we find that:
\begin{align}
    \frac{dy}{ds} &= 2 v_3+\chi'(s)+\psi'(s) \nonumber \\
    &= 2 v_3 + \frac{1}{2}\left[\tanh(\rho)+\coth(\rho)\right](v_1 \cos y-v_2 \sin y)-v_3 \nonumber \\
    &= v_3+ \frac{1}{2}\left[\frac{\sinh \rho}{\cosh \rho}+\frac{\cosh \rho}{\sinh \rho}\right](v_1 \cos y-v_2 \sin y) \nonumber \\
    &= v_3 + \coth(2\rho)(v_1 \cos y-v_2 \sin y).
\end{align}
 
We define $2\rho(s) =: x(s)$, so we find a system of two differential equations:
\begin{align}
    \frac{dy}{ds} &= v_3 + \coth(x)(v_1 \cos y-v_2 \sin y), \label{dyds}  \\
    \frac{dx}{ds} &= v_1 \sin(y)+ v_2 \cos(y), \label{dxds}
\end{align}
from which we get:
\begin{align}
\label{differentialdydx}
    \frac{dy}{dx} = \frac{v_3+ \coth(x)(v_1 \cos y - v_2\sin y)}{v_1 \sin(y)+v_2 \cos(y)}.
\end{align}

This equation is of the form $\frac{dy}{dx}= f(x,y)$ or equivalently, it can be expressed as:
\begin{align}
    M(x,y) dx + N(x,y) dy=0,
\end{align}
where $M(x,y)= -v_3- \coth(x)(v_1 \cos y-v_2 \sin y) $ and $N(x,y)= v_1 \sin(y)+v_2 \cos(y)$. For the equation $M dx+ Ndy=0$ to be an exact differential, we must 
satisfy the condition:
\begin{align}
    \frac{\partial M}{\partial y}= \frac{\partial N}{\partial x}.
\end{align}

For the differential equation \ref{differentialdydx}, we have:
\begin{align}
    \frac{\partial M}{\partial y}= \coth (x) (v_1 \sin (y)+v_2 \cos (y)), ~~~~ \frac{\partial N}{\partial x}=0,
\end{align}
which clearly shows that the differential equation \ref{differentialdydx} is not an exact differential equation. To solve the differential equation, we first have to make it exact, which is done by the integrating factor $\mu(x,y)$:
\begin{align}
    \mu(x,y) M(x,y) dx+ \mu(x,y) N(x,y) dy=0.
\end{align}

Let us, for the moment, assume that the integrating factor is just a function of $x$:
\begin{align}
    \mu(x) M(x,y) dx+ \mu(x) N(x,y) dy=0.
\end{align}

For the above equation, to be exact, we must have:
\begin{align}
    \frac{\partial}{\partial y}\left( \mu(x)M(x,y)\right) &= \frac{\partial}{\partial x}(\mu(x)N(x,y)), \nonumber  \\
    \mu \frac{\partial M}{\partial y} &= \frac{\partial \mu}{\partial x} N(x,y)+ \mu \frac{\partial N}{\partial x}, \nonumber \\
    \mu \bigg(\frac{\partial M}{\partial y}-\frac{\partial N}{\partial x}\bigg) &= \frac{d\mu}{dx}N(x,y), \nonumber\\
    \frac{d\mu}{\mu} &= \frac{\bigg(\frac{\partial M}{\partial y}-\frac{\partial N}{\partial x}\bigg)}{N(x,y)}dx.
\end{align}
For the differential equation under consideration, we can write the above equation as:
\begin{align}
    \frac{d\mu}{\mu} &= \coth(x) dx,
\end{align}
integration of which leads to the following solution:
\begin{align}
    \log(\mu) &= \log(\sinh(x))+\log C, \nonumber \\
    \mu &= C \sinh(x).
\end{align}

Using this integrating factor, we can rewrite the differential equation as:
\begin{align}
\label{diffFG}
   \underbrace{- \sinh(x)(v_3 + \coth(x)(v_1\cos y-v_2 \sin y))}_{G(x,y)} dx+ \underbrace{ \sinh(x)(v_1 \sin(y)+v_2\cos y)}_{F(x,y)} dy= 0,
\end{align}
where we got rid of the irrelevant multiplicative factor. 
Let us now verify that:
\begin{align}
    \frac{\partial G}{\partial y} = \cosh (x) (v_1 \sin (y)+v_2 \cos (y)) =  \frac{\partial F}{\partial x},
\end{align}
which shows that the differential corresponds to a total derivative of 
some function $H(x,y)$. In consequence, the following total derivative holds:
\begin{align}
dH = \frac{\partial H}{\partial x} dx +\frac{\partial H}{\partial y} dy = 0,
\end{align}
so that:
\begin{align}
    \frac{\partial H}{\partial x} &= G(x,y) =  - \sinh(x)v_3 -\cosh(x)(v_1\cos y-v_2 \sin y)), \\
    \frac{\partial H}{\partial y} &= F(x,y) = \sinh(x)(v_1 \sin(y)+v_2\cos y).
\end{align}
The above equations can be integrated as:
\begin{align}
    H(x,y) &=  - \cosh(x)v_3 -\sinh(x)(v_1\cos y-v_2 \sin y))+ h(y), \\
    H(x,y) &=   \sinh(x)(-v_1 \cos(y)+v_2\sin y) + g(x),
\end{align}
so that: 
\begin{equation}
g(x) = - \cosh(x)v_3 + C_1  \ \  \text{and}\ \ h(y)=C_2.
\end{equation}
In consequence, the solution can be written as:
\begin{equation}
\sinh(x)(v_1\cos y-v_2 \sin y)+\cosh(x)v_3 = C, 
\end{equation}
with some constant $C$. One can, at this stage, fix the 
the integration constant $C$ by considering the boundary condition 
at $s=0$, where:
\begin{equation}
x(0)=0 \ \ \text{and} \ \  y(0) = \psi(0)+\chi(0), 
\end{equation}
so that $U(s=0) = \mathbb{I}$. This implies that $C=v_3$.
Taking this into account, the solution can be written as:
\begin{equation}
(K^2-v_3^2)\cosh^2(x)+2v_3^2\cosh(x)-(v_3^2+K^2)=0.
\end{equation}
where $K:= v_1\cos y-v_2 \sin y$, so that $\Delta = 4K^4 \geq 0$. 
In consequence, there are two solutions for $\cosh(x)$:
\begin{equation}
\cosh(x) =\frac{ \pm K^2 -v_3^2}{K^2-v_3^2}.
\end{equation}
The ``+'' solution is trivial since it corresponds to the case $\cosh(x)=1$, 
satisfied for the initial value $x(s=0)=0$. The second (``-'') solution reads:
\begin{equation}
\label{eqncosh}
\cosh(x) = \frac{v_3^2+K^2}{v_3^2-K^2},
\end{equation}
which is valid for $v_3^2>K^2$. From here:
\begin{equation}
\label{eqntanhxby2}
K^2 = v_3^2 \tanh^2(x/2).
\end{equation}
Furthermore:
\begin{equation}
\label{Eqncothx}
\coth(x) = \pm \frac{v_3^2+K^2}{2v_3 K}.
\end{equation}

Let us now implement the boundary conditions:
We have $x(0)=0$ and $y(0)= \psi(0)+\chi(0)$. Substituting it in \ref{eqncosh}, we get:
\begin{align}
    1= \frac{v_3^2+\{v_1 \cos(\psi(0)+\chi(0))-v_2 \sin(\psi(0)+\chi(0))\}^2}{v_3^2-\{v_1 \cos(\psi(0)+\chi(0))-v_2 \sin(\psi(0)+\chi(0))\}^2}.
\end{align}
This gives us:
\begin{align}
\label{v1intermsofv2}
    \tan(\psi(0)+\chi(0))= \frac{v_1}{v_2}.
\end{align}

Similarly, at $s=1$, we have:
\begin{align}
    x(1)= 2\rho(1),~~~~ y(1)= 2 v_3+\chi(1)+\psi(1).
\end{align}

From Eq.\ref{eqncosh} we find: 
\begin{align}
    \cosh(2\rho(1)) &= \frac{v_3^2+ \{v_1\cos(y(1))-v_2 \sin(y(1))\}^2}{v_3^2- \{v_1\cos(y(1))-v_2 \sin(y(1))\}^2},
\end{align}
and using Eq. \ref{v1intermsofv2}, we can write the above equation as:
\begin{align}
    \cosh(2\rho(1))= \frac{v_3^2+ v_2^2[\tan(\psi(0)+\chi(0))\cos(y(1))- \sin(y(1))]^2}{v_3^2- v_2^2[\tan(\psi(0)+\chi(0))\cos(y(1))-\sin(y(1))]^2}.
\end{align}

Let us define $R:=\tan(\psi(0)+\chi(0))\cos(y(1))- \sin(y(1))$, so that:
\begin{align}
    \cosh(2\rho(1)) = \frac{v_3^2+v_2^2 R^2}{v_3^2-v_2^2R^2}.
\end{align}
This allows us to write the following relation:
\begin{align}
    v_2^2= \frac{v_3^2}{R^2}\tanh^2(\rho(1)).
\end{align}

With the relation between the $v_i$'s, the quantity $v_1^2+v_2^2+v_3^2$ can be written as:
\begin{align}
    v_1^2+v_2^2+v_3^2 &= v_2^2 \tan^2(\psi(0)+\chi(0))+ v_2^2+ v_3^2 \nonumber\\
    &= v_3^2\bigg(1+ \frac{1+\tan^2(\psi(0)+\chi(0))}{R^2}\tanh^2 \rho(1)\bigg).
\end{align}

The above expression can be simplified further by observing that for $U(s=0)=\mathbb{I}$, $\chi(0)$ must be $2n\pi$. This gives:
\begin{align}
    v_1^2+v_2^2+v_3^2= v_3^2\bigg(1+ \frac{1+\tan^2(\psi(0))}{R^2}\tanh^2 \rho(1)\bigg).
\end{align}

Therefore, we have:
\begin{align}
    C[U(\mathfrak{su(1,1)})]= |v_3|\sqrt{1+\tanh^2(\rho(1))\csc^2(y(1)-\psi(0))}.
\end{align}
Substituting $y(1)= 2 v_3+\chi(1)+\psi(1)$, in the above equation, we get:
\begin{align}
\label{complexitygeneral}
    C[U(\mathfrak{su(1,1)})]= |v_3|\sqrt{1+\tanh^2(\rho(1))\csc^2(2v_3+\chi(1)+\psi(1)-\psi(0))}
\end{align}
The above expression shows that the complexity of a unitary transformation belonging to the $SU(1,1)$ group can be entirely determined from $v_3$. Therefore, the task now is to determine $v_3$.

To determine $v_3$, we can use the differential equations 
for $\chi$ and $\psi$:
\begin{align}
    \chi'(s) &= \frac{1}{2}\left(\tanh(\rho(s))K-v_3\right), \\
    \psi'(s) &= \frac{1}{2}\left(\coth(\rho(s))K-v_3\right).
\end{align}
From Eq. \ref{eqntanhxby2}, we have:
\begin{align}
    \tanh(\rho(s))= \pm \frac{K}{v_3}.
\end{align}
Using this relation, the differential equations for $\psi$ and $\chi$ can be written as:
\begin{align}
    \chi'(s) &= \frac{1}{2}\left(\pm \frac{K^2}{v_3}-v_3\right),\\
    \psi'(s) &= \frac{1}{2}\left(\pm v_3-v_3\right).
\end{align}
The equation for $\chi$ is difficult to solve analytically. But $\psi$ can be solved:
\begin{align}
    \psi'(s)= \begin{cases}
        0 ~~ {\text {for the +ve solution}}\\
        -v_3 ~~ {\text {for the -ve solution}}
    \end{cases}~~.
\end{align}
From the above equation, we get the hint that there is a way of obtaining $v_3$ in terms of $\psi$.
The solution to the above equation can be written as:
\begin{align}
    \psi(s)= \begin{cases}
        \psi(s) &= v ~~ \text{for the +ve solution}\\
        \psi(s) &= -v_3 s+ v ~~~ \text{for the -ve solution}
    \end{cases}~~.
\end{align}
The positive solution is not useful since we cannot extract $v_3$ from it. However, the negative solution can be used. Implementing the boundary condition, we get:
\begin{align}
\label{eqnv3}
    v_3= \psi(0)-\psi(1).
\end{align}
This shows that $v_3$ can be entirely determined from the boundary values of the angular variable $\psi$. The values of $\chi$ are not even required. Eq. \ref{eqnv3} shows that $v_3$ is the difference between two angular variables. One can always argue that there is an ambiguity in the value of $v_3$, and when determined from only $\psi$, it doesn't contain the entire information. There is always a possibility that $v_3$, when determined from $\chi(s)$, might give a shorter path. But since both $\chi$ and $\psi$ are angular quantities, their values are bounded.

From the generic element written in \ref{genericelement}, it is obvious that the angles $\psi$ and $\chi$ lie between $[0,2\pi)$. This essentially means that the quantities $\psi(0)$ and $\psi(1)$ appearing in the expression of complexity can only take values between 0 and $2\pi$. Therefore, we have:
\begin{align}
    \chi(0/1)= \chi(0/1)~{\rm mod}~2\pi, ~~ \psi(0/1)= \psi(0/1)~{\rm mod}~2\pi.
\end{align}
$v_3$ is the difference between the angular values $\psi(0)$ and $\psi(1)$ i.e.
\begin{align}
    v_3= \psi(0)-\psi(1) ~~\in~~ (-2\pi, 2\pi),
\end{align}
which implies:
\begin{align}
    |v_3|= |\psi(0)-\psi(1)| ~ \in~[0,2\pi).
\end{align}
$|v_3|$ represents the angular difference between the values of $\psi$ at $s=0$ and $s=1$. The definition of complexity requires choosing the minimum value of $v_3$. Hence, we have:
\begin{align}
    |v_3|= {\rm min}\big(|\psi(0)-\psi(1)|,~ 2\pi- | (\psi(0)-\psi(1)|\big).
\end{align}

Having derived the general expression of complexity for any unitary operator written in terms of the $\mathfrak{su}(1,1)$ generators, we will consider some explicit examples that are useful for our purpose, namely the rotation operator, the squeezing operator, and the product of these two.

\subsection{Rotation operator complexity}

The two-mode rotation operator can be written as:
\begin{align}
    \hat{R}_2(\theta) &= \exp\bigg\{i \theta_k(\hat{c}^{\dagger}_{\bf k}\hat{c}_{\bf k}+ \hat{c}^{\dagger}_{-{\bf k}}\hat{c}_{-{\bf k}}+1)\bigg\},
\end{align}
which, in terms of $O_i$'s, can be written as:
\begin{align}
    \hat{R}_2(\theta)= \exp (2i\theta_k \hat{\mathcal{O}}_3).
\end{align}
Using the matrix representation of the generators, this can be written as:
\begin{align}
    \begin{pmatrix}
        e^{i\theta_k} & 0\\ \\
        0 & e^{-i\theta_k}       
    \end{pmatrix}.
\end{align}

Comparing the above unitary with the generic element Eq. \ref{genericelement}, we get:
\begin{align}
\label{boundary1}
    \rho(1)= 0, ~~ \chi(1)= \theta_k, ~~ \psi(1)= {\rm unspecified}
\end{align}
Similarly, when $s=0$, which is the identity operator, we get the following condition:
\begin{align}
\label{boundary2}
    \rho(0)=0,~~ \chi(0)=0,~~ \psi(0)= {\rm unspecified}
\end{align}

In the derivation of the general case, we found that $v_3$ can be determined from the equation from the boundary values of the angular quantity $\psi$ from the following equation:
\begin{align}
    v_3= \psi(0)-\psi(1).
\end{align}
However, from the boundary values written in Eq. \ref{boundary1} and \ref{boundary2}, we notice that $\psi(0)$ and $\psi(1)$ are unspecified quantities, which means that $v_3$ cannot be determined from $\psi$. So the only option left is to solve the equation for $\chi$ and use it to determine $v_3$.
The equation for $\chi$ is given by:
\begin{align}
    \chi'(s)= \frac{1}{2}\bigg(\pm \frac{K^2}{v_3^2}-v_3\bigg).
\end{align}

From Eq. \ref{eqntanhxby2}, we have:
\begin{align}
    K^2= v_3^2 \tanh^2(\rho(s)),
\end{align}

where $K= v_1 \cos(y)-v_2 \sin(y)$. Imposing the boundary condition in the above equation, we have:
\begin{align}
    \{v_1 \cos(y(0))-v_2 \sin(y(0))\}^2 &= v_3^2 \tanh^2(\rho(0)),\\
    \{v_1 \cos(y(1))-v_2 \sin(y(1))\}^2 &= v_3^2 \tanh^2(\rho(1)).
\end{align}

Since, $\rho(0)=0$ and $\rho(1)=0$, the above equation gives:
\begin{align}
    v_1=0 ,~~ v_2=0.
\end{align}

With $v_1=v_2=0$, the differential equation for $\chi$ reduces to:
\begin{align}
    \chi'(s)= -\frac{v_3}{2},
\end{align}
integration of which gives: 
\begin{align}
    \chi(s)= -\frac{v_3}{2}s+ v.
\end{align}

Implementing the condition $\chi(0)=0$, the constant $v=0$, which gives:
\begin{align}
    \chi(s)= -\frac{v_3}{2}s.
\end{align}
In consequence, $v_3$ can be written as:
\begin{align}
    v_3= -2\chi(1)= -2\theta_k.
\end{align}

The complexity, therefore, is given by:
\begin{align}
    C[R_2(\theta)]& = \min |v_3|= 2 \min \left[|\chi(1)|\mod 2\pi, |2\pi-\chi(1)|\mod 2\pi  \right] \nonumber \\
                     & = 2 \min \left[ |\theta| \mod 2\pi, |2\pi-\theta| \mod 2\pi   \right].
\end{align} 

\subsection{Squeezing operator complexity}
\label{sec3}
The two-mode squeezing operator can be written as:
\begin{align}
    S_2(r,\phi) &= \exp\bigg\{r_k (e^{-i\phi}\hat{c}_{\Vec{k}}\hat{c}_{-\Vec{k}}-e^{i\phi}\hat{c}^{\dagger}_{\Vec{k}}\hat{c}^{\dagger}_{-\Vec{k}})\bigg\}.
\end{align}

In terms of the $\mathfrak{su}(1,1)$ generators $O_I$'s, it can be written as: 
\begin{align}
    S_2(r,\phi) &= \exp\bigg[r_k(e^{-i\phi_k}(\mathcal{O}_1-i\mathcal{O}_2)- e^{i\phi_k}(\mathcal{O}_1+i\mathcal{O}_2))], \\
    &= \exp\bigg[-2ir_k (\sin(\phi_k)\mathcal{O}_1+ \cos(\phi_k)\mathcal{O}_2)\bigg].
\end{align}

Using the matrix representation of the generators it can be expressed as: 
\begin{align}
\label{squeezedoperator}
S_2(r,\phi)=
    \begin{pmatrix}
        \cosh(r_k) & -e^{i\phi_k}\sinh(r_k)\\ \\
        -e^{-i\phi_k}\sinh(r_k) & \cosh(r_k)        
    \end{pmatrix}.
\end{align}

Using the boundary condition $U(s=1)=S_2(r,\phi)$, we get:
\begin{align}
    \rho(1)= r_k, ~~ \chi(1)= 2n\pi, ~~~ \psi(1)= \phi_k.
\end{align}

Similarly, from the condition $U(s=0)= \mathbb{I}$, we obtain:
\begin{align}
    \rho(0)=0, ~~ \chi(0)= 2k\pi,~~~ \psi(0)= {\rm unspecified}.
\end{align}

Based on the above, we find: 
\begin{align}
    y(1)= 2 v_3+\chi(1)+\psi(1) =2 v_3+ 2n\pi+\phi_k.
\end{align}

From the general expression of the complexity of any operator belonging to $SU(1,1)$:
\begin{align}
    C[U(\mathfrak{su}(1,1))]= |v_3|\sqrt{1+\tanh^2(\rho(1))\csc^2(2v_3+\chi(1)+\psi(1)-\psi(0))},
\end{align}
we now find: 
\begin{align}
    C[S_2(r_k,\phi_k)] &= |v_3|\sqrt{1+\tanh^2(r_k)\csc^2(2v_3+2n\pi+ \phi_k-\psi(0))} \nonumber \\
    &= |v_3|\sqrt{1+\tanh^2(r_k)\csc^2(2v_3+ \phi_k-\psi(0))}.
\end{align}

Because $v_3$ can be determined from:
\begin{align}
    v_3= \psi(0)-\psi(1),
\end{align}
we obtain the following expression on the complexity of the two-mode squeezing operator: 
\begin{align}
    C[S_2(r_k,\phi_k)] & =({\rm min}|\phi_k-\psi(0)|)\sqrt{1+\tanh^2(r_k)\csc^2(\phi_k-\psi(0))}.
    \label{ComplexityS2psi0}
\end{align}

The quantity $\psi(0)$ is an unspecified quantity. This is because 
the parametrization \ref{genericelement} leads to $U(0)=\mathbb{I}$, 
irrespectively on the value of $\psi(0)$. Therefore, $v_3$ is determined 
up to the unspecified quantity.

In the limit $r \rightarrow 0$, $S_2(r_k,\phi_k)$ written in 
\ref{squeezedoperator} tends to $\mathbb{I}$. However, there 
may still be some phase difference between $\phi$ and $\psi(0)$. 
It is up to the definition of the complexity whether or not to take  
this factor into account. 

For the complexity to vanish in this limit, we must have $\psi(0)\rightarrow \phi_k$.
For $\psi(0) \rightarrow \phi_k $, the complexity for the target unitary operator can be written as:
\begin{align}
\label{complexitysqueezedfinal}
    C[U_{\rm target}] \approx |\tanh(r_k)|.
\end{align} 

Furthermore, such a choice guarantees that the expression 
(\ref{ComplexityS2psi0}) does not diverge at $\phi_k-\psi(0) = 
\pi$ due to the $\csc(\phi_k-\psi(0))$ term. Such divergence 
could be considered unphysical. 

We must note that in the above derivation, the squeezing amplitude 
and angle are treated as independent parameters. If we consider time-dependent
parameters, then fixing the unspecified quantity requires additional consideration. 

For time dependent $r_k$ and $\phi_k$, we get:
\begin{align}
    C[S_2(r(t),\phi(t))]= ({\rm min}|\phi_k(t)-\psi(0)|)\sqrt{\tanh ^2(r_k(t)) \csc ^2(\phi_k(t) -\psi(0) )+1}.
\end{align}
At some initial time $t_0$, when the squeezing operator tends to identity, we have $r_k( t_0)\rightarrow 0$ and $\phi_k(t_0)$. This gives us:
\begin{align}
    \psi(0)= \phi_k(t_0).
\end{align}
Substituting this, we get the complexity of the time dependent squeezing operator as:
\begin{align}
    C[S_2(r(t),\phi(t))]= ({\rm min}|\phi_k(t)-\phi(t_0)|)\sqrt{\tanh ^2(r_k(t)) \csc ^2(\phi_k(t) -\phi(t_0) )+1}.
\end{align}

Let us notice that this equation faces divergence at $\phi_k(t) -\phi(t_0)=\pi$. 

\subsection{Evolution operator complexity}
\label{sec:complexityTE}

The dynamics of the system are captured by the product of the squeezing and the rotation operator, which is the target unitary operator in this scenario.
\begin{align}
\label{evolutionoperator}
    U_{\rm evolution}= S_2(r(t),\phi(t))R_2(\theta(t)).
\end{align}

In \cite{Chowdhury:2024ntx}, the complexity upper bound was derived using a representation-independent approach within a certain approximation in the operator expansion, which is as follows:
\begin{align}
\label{upperbound}
C[U_{\rm target}^{(1)}] \lesssim 2  \sqrt{\theta(t)^2(1+4~r(t)^2 \csc ^2(2 \theta(t) ))},
\end{align}
and improved version of the above formula was also derived:
\begin{align}
\label{improvedupperbound}
    C[U_{\rm target}^{(2)}] \lesssim 2\sqrt{\theta(t)^2(1+4r(t)^2(1+\theta(t)^2)\csc^2(2\theta(t)))}\,.
\end{align}

As discussed in the previous subsection, using a finite-dimensional matrix representation of the generators, an exact formula for the complexity of the time evolution operator can be derived. The consideration of the matrix representation of the generators gives us the precise position of the target unitary operator in the unitary group manifold. The target unitary operator written in Eq. \ref{evolutionoperator} is a product of two exponential operators. In the representation-independent approach, to accurately implement the boundary condition, it was necessary to convert the product of exponentials into a single exponential, which can be done by the BCH formula. The analysis carried out in \cite{Chowdhury:2024ntx}, considered only a few terms in the BCH formula which effectively changed the position of the target unitary operator in the unitary group manifold. The consideration of a finite-dimensional matrix representation of the $\mathfrak{su}(1,1)$ generators overcomes this problem and allows us to incorporate the boundary condition in terms of finite-dimensional matrices accurately.

Using the matrix representation of the $O_I$'s, the target unitary operator written in Eq. \ref{evolutionoperator} can be expressed as:
\begin{align}
\label{evolutionoperatormatrix}
U_{\rm evolution}=
    \begin{pmatrix}
        e^{i\theta_k(t)}\cosh(r_k(t)) & -e^{-i(\theta_k(t)-\phi_k(t))}\sinh(r_k(t))\\ \\
        -e^{i(\theta_k(t)-\phi_k(t))}\sinh(r_k(t)) & e^{-i\theta_k(t)}\cosh(r_k(t))        
    \end{pmatrix}.
\end{align}

Comparing $U_{\rm evolution}$ with the generic element, 
\begin{align}
    U(s)= \begin{pmatrix}
        \cosh(\rho(s))e^{i\chi(s)} && -\sinh(\rho(s))e^{i\psi(s)} \\
        -\sinh(\rho(s))e^{-i\psi(s)} && \cosh(\rho(s))e^{-i\chi(s)}
    \end{pmatrix},
\end{align}
 at $s=1$, we get: 
\begin{align}
\label{parameterrelation}
    \rho(1)= r_k(t), ~~ \psi(1)= \phi_k(t)-\theta_k(t), ~~ \chi(1)= \theta_k(t).
\end{align}

Similarly, using the condition $U(s=0)= \mathbb{I}$, we get:
\begin{align}
    \rho(0)= 0, ~~ \chi(0)= 2n\pi, ~~ \psi(0)= {\rm unspecified}.
\end{align}

In this case again, $v_3$ can be determined from the values of 
$\psi$ at $s=0$ and $s=1$, so that:
\begin{align}
    v_3= \psi(0)-\psi(1),
\end{align}
which leads to the following equation for the geometric complexity:
\begin{align}
    C[U(\mathfrak{su(1,1)})]= |v_3|\sqrt{1+\tanh^2(\rho(1))\csc^2(2v_3+\chi(1)+\psi(1)-\psi(0))}.
\end{align}
Substituting the boundary values, we get:
\begin{align}
    v_3= \psi(0)-\phi_k(t)+\theta_k(t),
\end{align}
using which the expression of complexity for the evolution operator can be written as:
\begin{align}
\label{eqncomplexityevolution}
    C[U_{\rm evolution}] &= ({\rm min}|\psi(0)-\phi_k(t)+\theta_k(t)|) 
    \nonumber\\
   &\times  \sqrt{1+\tanh^2(r_k(t))\csc^2(\psi(0)-\phi_k(t)+2\theta_k(t))}.
\end{align}

The unspecified quantity $\psi(0)$ can be determined from the requirement that $C[U_{\rm evolution}]$ goes to 0 when the evolution operator tends to identity. 

Let us consider that in the limit $t\rightarrow -\infty$ (some initial time), $U_{\rm evolution} \rightarrow \mathbb{I}$. Therefore, we have:
\begin{align}
    & r_k(t\rightarrow -\infty) \rightarrow 0, \\
    \label{thetacondition1}
    & \theta_k(t \rightarrow -\infty) \rightarrow 2m \pi, ~~ m ~\epsilon ~\mathbb{N}.
\end{align}
In this limit, the expression of complexity can be rewritten as:
\begin{align}
    C[U]\bigg|_{t\rightarrow -\infty} &= |\psi(0)-\phi_k(-\infty)+\theta_k(-\infty)|
   \nonumber \\
   &\times \sqrt{1+\tanh^2(r_k(-\infty))\csc^2(2\theta_k(-\infty)+\psi(0)-\phi_k(-\infty))}.
\end{align}
For $C[U]\bigg|_{t\rightarrow -\infty}$ to vanish in this limit, we must have:
\begin{align}
\label{thetacondition2}
    \psi(0) = \phi_k(-\infty)-\theta_k(-\infty), ~~~ \text{and} ~~~ \theta_k(-\infty)= n \pi.
\end{align}
For \ref{thetacondition1} and \ref{thetacondition2} to hold simultaneously, the following 
must hold:
\begin{align}
    n= 2m.
\end{align}
Therefore, using the above condition, we get:
\begin{align}
\label{psi0evolution}
    \psi(0)= \phi_k(-\infty)-2m\pi.
\end{align}
Putting all these conditions together, we find:
\begin{align}
\label{complexitytimeevolutionoperator}
    C[U_{\rm evolution}] &= ({\rm min}|\phi_k(-\infty)-\phi_k(t)+\theta_k(t)-2m \pi|)\nonumber  \\
                        &\times \sqrt{1+\tanh^2(r_k(t))\csc^2(2\theta_k(t)-\phi_k(t)+\phi_k(-\infty)-2m\pi)}.
\end{align}

Let us now check the limiting conditions, where the target unitary operator reduces to just the squeezing and the rotation operator separately. In the limit, $\theta_k(t) \rightarrow 0$, we have:
\begin{align}
    U_{\rm evolution} \rightarrow S_2(r(t),\phi(t)),
\end{align}
and Eq. \ref{complexitytimeevolutionoperator} becomes:
\begin{align}
    C[U_{\rm evolution}(\theta_k(t)\rightarrow 0)] = ({\rm min}|\phi(t_0)-\phi_k(t)|)\sqrt{1+\tanh^2(r_k(t))\csc^2(\phi(t_0)-\phi_k(t))}
\end{align}
One must note that the value of $m$ should be 0 in this case as $\theta_k \rightarrow 0$ for every $t$. This shows that the complexity obtained for the product of the squeezing and the rotation operator reduces to the complexity of the squeezing operator in the limit where the rotation operator goes to identity. 
Similarly, the evolution operator reduces to just the rotation operator in the limit $r_k(t) \rightarrow 0$ irrespectively of the value of $\phi_k(t)$. This means that $\phi_k(t)$ is an unspecified quantity in this case. With these considerations, Eq. \ref{eqncomplexityevolution}, can be written as:
\begin{align}
    C[U_{\rm evolution}] &= {\rm min}(|\psi(0)-\phi_k(t)+\theta_k(t)|) = {\rm min}(A+\theta_k(t)),
    \\~\nonumber &~~~~~~~~~~A= \psi(0)-\phi_k(t) \rightarrow {\rm unspecified}. 
\end{align} 
In the above expression, the quantity $\psi(0)-\phi_k(t)$ is unspecified. This arises because of the ambiguity of determining $v_3$. To match the result with what we obtained for the rotation operator, the quantity $A$ must be equal to $\theta_k(t)$.

\section{Relative complexity of two evolution operators}
\label{sec4}

In the previous section, we determined the complexity of the product
of the squeezing and rotation operators. Geometrically, this
corresponds to finding the shortest geodesic on the $SU(1,1)$ group
manifold connecting the identity operator to the target operator. 
We now extend this framework by introducing the notion of
\textit{relative complexity}, which measures the complexity of
reaching a given operator starting from a nontrivial reference
operator rather than from the identity. In this setting, we are
interested in the relative complexity between two operators, each 
given by a product of squeezing and rotation operators, but
characterized by different values of the squeezing and rotation
parameters.

The relative complexity of two operator $\hat{U}_1$ and $\hat{U}_2$ 
is denoted by $C[\hat{U}_1,\hat{U}_2]$. Due to the right-invariance 
of complexity, it can be shown that the relative between two 
operators $\hat{U}_1$ and $\hat{U}_2$ is equal to the complexity of 
the operator $\hat{U}_1\hat{U}_2^{-1}$. Therefore, we have:
\begin{equation}
C(\hat{U}_1,\hat{U}_2) = C(\hat{U}_1\hat{U}_2^{-1},\hat{\mathbb{I}}).
\end{equation}

The matrix elements of $\hat{U}_1\hat{U}_2^{-1}=\hat{U}$ can be written as:
\begin{align}
    U^{11} &= e^{-i(\theta_2-\theta_1)}\cosh(r_1)\cosh(r_2)-e^{-i(\theta_1-\theta_2+\phi_2-\phi_1)}\sinh(r_1)\sinh(r_2),\\
    U^{22} &= e^{i(\theta_2-\theta_1)}\cosh(r_1)\cosh(r_2)-e^{i(\theta_1-\theta_2+\phi_2-\phi_1)}\sinh(r_1)\sinh(r_2),\\
    U^{12} &= -e^{-i(\theta_1-\theta_2-\phi_1)}\cosh(r_2)\sinh(r_1)+e^{-i(\theta_2-\theta_1-\phi_2)}\cosh(r_1)\sinh(r_2),\\
    U^{21} &= -e^{i(\theta_1-\theta_2-\phi_1)}\cosh(r_2)\sinh(r_1)+e^{i(\theta_2-\theta_1-\phi_2)}\cosh(r_1)\sinh(r_2).
\end{align}

As a simplification, we consider the case where the squeezing angle $\phi$ and the rotation angle $\theta$ of both the unitary operators are taken to be zero. This means that the unitary operators $\hat{U}_1$ and $\hat{U}_2$ are just characterized by the squeezing parameters $r_1$ and $r_2$, respectively. For this simple choice, the target unitary operator now becomes:
\begin{align}
    U^{11}=U^{22}= \cosh(r_1-r_2), ~~~ U^{12}=U^{21}=\sinh(r_2-r_1).
\end{align}

Comparing with the general form of $\hat{U}$ at $s=1$, we get:
\begin{align}
    \rho(1)= r_1-r_2, ~~~~ \psi(1)= 2n\pi, ~~~~ \chi(1)= 2n\pi.
\end{align}
The relative complexity of $\hat{U}_1(r_1,0,0)$ and $\hat{U}_2(r_2,0,0)$ is found to be:
\begin{align}
   C(\hat{U}_1,\hat{U}_2)= |\psi(0)-2n\pi|\sqrt{1+\csc^2(\psi(0)-2n\pi)\tanh^2(r_1-r_2)}.
\end{align}

In the above formula, interchanging $r_1$ with $r_2$, keeps the formula 
unchanged proving that the property $C[\hat{U}_1,\hat{U}_2]= 
C[\hat{U}_2,\hat{U}_1]$. 

We now verify whether our formula satisfies the triangle 
inequality. To this end, we define:
\begin{align}
    \psi(0)-2n\pi =: \delta, ~~~~~~ r_i-r_j=: r_{ij},
\end{align}
which allows us to re-express $C[U_i,U_j]$ as:
\begin{align}
    C[\hat{U}_i,\hat{U}_j]= |\delta|\sqrt{1+\csc^2(\delta)\tanh^2(r_{ij})}.
    \label{RelativeComplexity1}
\end{align}

The undetermined quantity $\delta$ can be fixed from the requirement
that $C[\hat{U}_i,\hat{U}_i]=0$. In the limit $r_i\rightarrow r_j$, 
when the initial and the final operators are identical, the complexity
should tend to 0. To satisfy this condition, $\delta$ must tend to $0$
simplifies  (\ref{RelativeComplexity1}) to:
\begin{align}
    C[\hat{U}_i,\hat{U}_j]= |\tanh(r_{ij})|.
    \label{RelativeComplexity1}
\end{align}

To prove the triangle inequality, we need to show that:
\begin{align}
    C[\hat{U}_1,\hat{U}_2]\leq C[\hat{U}_1,\hat{U}_3]+C[\hat{U}_3,\hat{U}_2],
\end{align}
which can be written as:
\begin{align}
    |\tanh(x+y)| \leq |\tanh(x)|+ |\tanh(y)|,
    \label{triangleInequality2}
\end{align}
where $x=r_{13}$, $y=r_{32}$ and $x+y=r_{13}+r_{32}=r_{12}$.

To prove (\ref{triangleInequality2}), let us recall the 
standard triangle inequality:
\begin{equation}
|x+y| \leq |x|+|y|, 
\end{equation}
which holds for any $x,y \in \mathbb{R}$. Because 
$\tanh(x)$ is a monotonic function, so we can write:
\begin{equation}
\tanh(|x+y|) \leq \tanh(|x|+|y|). 
\end{equation}
Then let us notice that $\tanh(|x|)=|\tanh(x)|$, 
which allows us to rewrite the left-hand side of 
the inequality to the desired form:
\begin{equation}
|\tanh(x+y)| \leq \tanh(|x|+|y|). 
\end{equation}
Now, let us apply the hyperbolic identity:
\begin{equation}
\tanh(x + y) = \frac{\tanh(x) + \tanh(y)}{1 + \tanh(x)\tanh(y)},
\end{equation}
to the right-hand side of the inequality:
\begin{equation}
|\tanh(x+y)| \leq \frac{\tanh(|x|) + \tanh(|y|)}{1 + \tanh(|x|)\tanh(|y|)}. 
\end{equation}
Because the denominator satisfies the inequality:
\begin{equation}
1 + \tanh(|x|)\tanh(|y|) \geq 1, 
\end{equation}
we find: 
\begin{equation}
|\tanh(x+y)| \leq \frac{\tanh(|x|) + \tanh(|y|)}{1 + \tanh(|x|)\tanh(|y|)} \leq  \tanh(|x|) + \tanh(|y|).
\end{equation}
By applying the $\tanh(|x|)=|\tanh(x)|$ property to the 
right-hand side, the desired expression is obtained:
\begin{equation}
|\tanh(x+y)| \leq |\tanh(x)| + |\tanh(y)|,
\end{equation}
which completes the proof of the inequality 
(\ref{triangleInequality2}).

Thus, we verified the fact that the formula of relative 
complexity between two unitary operators $U_1(r_1,0,0)$ 
and $U_2(r_2,0,0)$ satisfies the triangle inequality. 

This indicates that the complexity formula obtained, 
at least for the fixed angles $\theta=0=\phi$, 
satisfies the four defining conditions of a metric:
\begin{itemize}
\item $C[\hat{U}_1,\hat{U}_1]=0$.
\item $C[\hat{U}_1,\hat{U}_2] >0$ for $\hat{U}_1\neq \hat{U}_2$ (Positivity).
\item $C[\hat{U}_1,\hat{U}_2]=C[\hat{U}_2,\hat{U}_1]$ (Symmetry). 
\item $C[\hat{U}_1,\hat{U}_2]\leq C[\hat{U}_1,\hat{U}_3]+C[\hat{U}_3,\hat{U}_2]$ 
(Triangle inequality).
\end{itemize}

The verification of these properties in the general case with 
$\theta \neq 0$ and $\phi \neq 0$ is technically demanding, and we leave a comprehensive investigation to future studies.

\section{Revisiting the harmonic oscillator and the inverted harmonic oscillator}

In this section, we will derive the complexity of the time evolution operators associated with the Hamiltonian of a harmonic oscillator and the inverted harmonic oscillator. The complexity of the time evolution operators for these systems using an upper bound approximation was studied in \cite{Chowdhury:2023iwg}. It is instructive to reinvestigate the complexity of the time evolution operators of these systems, going beyond the upper bound approximation. The purpose of this is to verify how far the upper bound result is from the exact complexity. Moreover, these systems serve as toy models for the scalar field on de Sitter spacetime. In the sub-Horizon limit, the scalar field mode behaves like a harmonic oscillator and like an inverted oscillator in the super-Horizon scales.

\subsection{The harmonic oscillator}
In terms of the creation and the annihilation operator, the harmonic oscillator Hamiltonian can be written as:
\begin{align}
    \hat{H}= \frac{\omega}{2} (\hat{c}^{\dagger}\hat{c}+ \hat{c} \hat{c}^{\dagger}).
\end{align}
The following hermitian operators can be used to construct the target unitary operator, $\hat{U}_{\rm target}= e^{-i\hat{H}t}$:
\begin{align}
    \hat{O}_1= \frac{\hat{c}^2+\hat{c}^{\dagger 2}}{4},~~  \hat{O}_2= \frac{i(\hat{c}^2-\hat{c}^{\dagger 2})}{4},~~ \hat{O}_3 = \frac{\hat{c}^{\dagger}\hat{c}+ \hat{c} \hat{c}^{\dagger}}{4}.
\end{align}

These operators satisfy the $\mathfrak{su}(1,1)$ Lie algebra.
In terms of these operators, the target unitary operator can be written as:
\begin{align}
    \hat{U}_{\rm target}= e^{-2i\omega t \hat{O}_3}. 
\end{align}

 Using the matrix representation of the operators $O_i$ written in \ref{matrixgenerators}, the target unitary operator can be written as:
\begin{align}
    U_{\rm target}= 
    \begin{pmatrix}
        e^{-i\omega t} & 0 \\
        0 & e^{i\omega t}
    \end{pmatrix}.
\end{align}

Comparing the above target unitary operator with the generic element: 
\begin{align}
    U(s)= \begin{pmatrix}
        \cosh(\rho(s))e^{i\chi(s)} && -\sinh(\rho(s))e^{i\psi(s)} \\
        -\sinh(\rho(s))e^{-i\psi(s)} && \cosh(\rho(s))e^{-i\chi(s)}
    \end{pmatrix}, 
\end{align}
we get:
\begin{align}
    \rho(1)= 0, ~~ \chi(1)= -\omega t, ~~~ \psi(1)= {\rm unspecified}.
\end{align}

At $t=0$, the target unitary operator is essentially the identity operator. We find: 
\begin{align}
    \rho(0)=0, ~~ \chi(0)= 0, ~~ \psi(0)= {\rm unspecified}.
\end{align}

This is equivalent to the rotation operator with the rotation angle characterized by $\omega t$. Hence, following the derivation of the rotation operator case, we get:
\begin{align}
    v_3= -2\chi(1).
\end{align}

So the complexity can be written as:
\begin{align}
    C[U_{\rm target}]& = \min |v_3|=  \min \left[ |2\chi(1)| \mod 2\pi, |2\pi-2\chi(1)| \mod 2\pi   \right] \nonumber \\
                     & =  \min \left[ |2\omega t| \mod 2\pi, |2\pi-2\omega t| \mod 2\pi   \right].
\end{align}

Therefore, the complexity of the time evolution operator of the harmonic oscillator exhibits the usual oscillatory behavior.

\subsection{Inverted harmonic oscillator}

In terms of the creation and annihilation operators, the inverted harmonic oscillator Hamiltonian can be written as:
\begin{align}
    \hat{H}_{\rm IHO}= -\frac{\Omega}{2}\left(\hat{c}^2+\hat{c}^{\dagger 2}\right).
\end{align}

The target unitary operator in this case can be written as:
\begin{align}
    U_{\rm IHO}= e^{-iH_{\rm IHO}t}= e^{2i\Omega t O_1}.
\end{align}

Using the matrix representation of the operators, the target unitary operator can be written as:
\begin{align}
    U_{\rm IHO}=\left(
\begin{array}{cc}
 \cosh (\Omega t ) & i \sinh (\Omega t) \\
 -i \sinh (\Omega t) & \cosh (\Omega t) \\
\end{array}
\right).
\end{align}

Comparing the above target unitary operator with the generic element: 
\begin{align}
    U(s)= \begin{pmatrix}
        \cosh(\rho(s))e^{i\chi(s)} && -\sinh(\rho(s))e^{i\psi(s)} \\
        -\sinh(\rho(s))e^{-i\psi(s)} && \cosh(\rho(s))e^{-i\chi(s)}
    \end{pmatrix}, 
\end{align}
we get at $s=1$:
\begin{align}
    \rho(1)= \Omega t,~~ \chi(1)= 0, ~~ \psi(1)= -\frac{\pi}{2}.
\end{align}

Similarly, for $s=0$, we find:
\begin{align}
    \rho(0)=0, ~~ \chi(0)=0, ~~ \psi(0)= {\rm unspecified}.
\end{align}

We can determine $v_3$ from the boundary values of $\psi$ as:
\begin{align}
    v_3= \psi(0)-\psi(1),
\end{align}
which gives the complexity as:
\begin{align}
    C[U_{IHO}] &= |\psi(0)-\psi(1)|\sqrt{1+\csc^2(\psi(1)-\psi(0))\tanh^2(\Omega t)}.
\end{align}

The unspecified quantity $\psi(0)$ can be determined from the condition that at $t=0$, when the target unitary operator tends to $\mathbb{I}$, the complexity should be 0. This means that the complexity of the target unitary operator can be written as:
\begin{align}
    C[U_{\rm IHO}]= |\tanh{\Omega t}|.
\end{align}

\begin{figure}
    \centering
    \includegraphics[width=10cm]{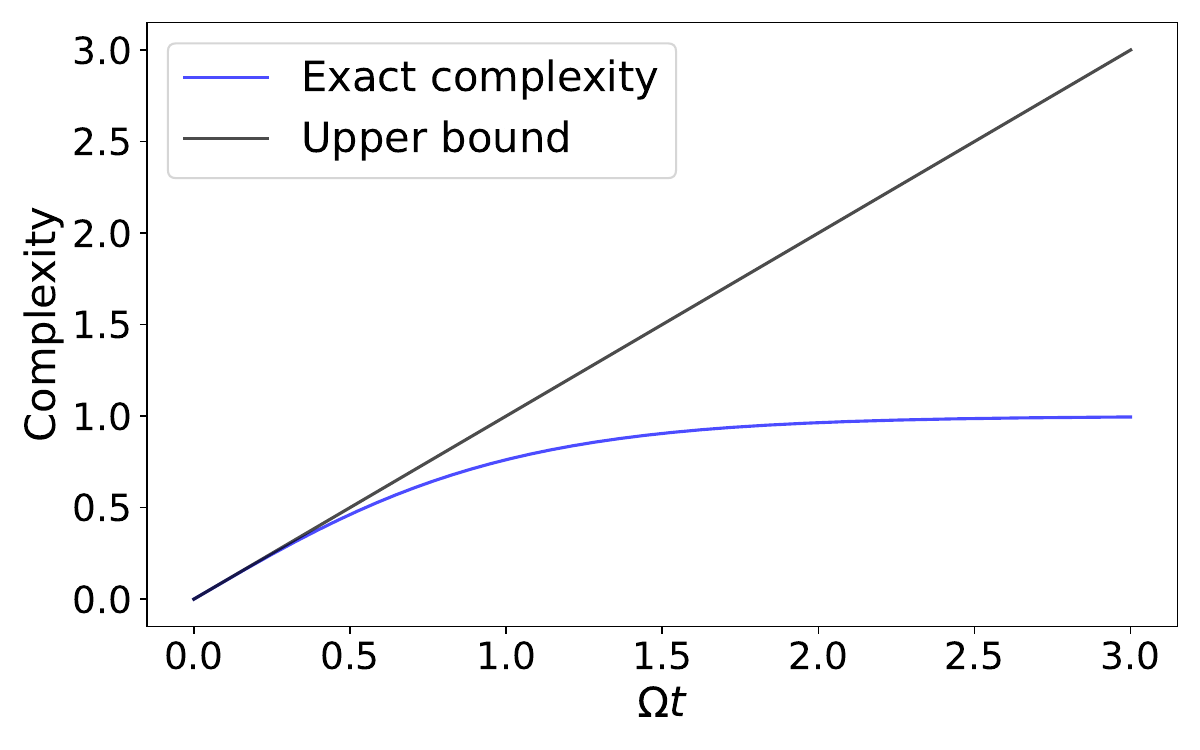}
    \caption{Complexity of the time evolution operator of an inverted harmonic oscillator. The upper bound is based on the formula 
    derived in Ref. \cite{Chowdhury:2024ntx}.}
    \label{fig:enter-label}
\end{figure}

One observes that, at small $t$, the exact complexity coincides 
with the upper-bound result derived in Ref.~\cite{Chowdhury:2024ntx}. 
However, the upper bound does not capture the saturation of the
complexity at late times.

\section{Application 1: de Sitter model}
\label{sec5}

In the previous section, we derived the formula for the exact complexity of the product of squeezing and rotation operators by utilizing a suitable finite-dimensional matrix representation of the generators of $\mathfrak{su}(1,1)$. We revisit the problem of complexity 
of the evolution of a massless test scalar field in the de Sitter background studied in \cite{Chowdhury:2024ntx}. The time-dependent effective frequency for a scalar field 
minimally coupled to a gravitationally background is given by:
\begin{align}
    \omega(\tau)^2= k^2+m^2 a(\tau)^2-\frac{a''}{a},
\end{align}
which, for a massless scalar field on a de Sitter background, simplifies to:
\begin{align}
\label{desitterfrequency}
    \omega^2_{dS}(\tau)= k^2-\frac{2}{\tau^2}.
\end{align}
With this frequency profile, the mode function satisfies the equation:
\begin{align}
    f_k''(\tau)+\bigg(k^2-\frac{2}{\tau^2}\bigg)f_k(\tau)=0,
\end{align}
having the general solution \cite{Mukhanov:2007zz}:
\begin{align}
    f_k= -A_k\bigg(1-\frac{i}{k\tau}\bigg)\frac{e^{-ik\tau}}{\sqrt{2k}}-B_k\bigg(1+\frac{i}{k\tau}\bigg)\frac{e^{ik\tau}}{\sqrt{2k}}.
\end{align}

The constants $A_k$ and $B_k$ determine the mode functions and must be chosen 
appropriately to obtain a physically motivated vacuum state. For de Sitter, 
the preferred vacuum state is the \textit{Bunch-Davies} vacuum, which is 
essentially the Minkowski vacuum in the early time limit, i.e., 
$\tau \rightarrow -\infty$. Therefore, the mode function is given by:
\begin{align}
    f_k(\tau)= \frac{e^{-ik\tau}}{\sqrt{2k}}\bigg(1-\frac{i}{k\tau}\bigg).
\end{align}

Since the mode functions are explicitly available in this case, the
time-dependent Bogoliubov coefficients capturing the evolution 
can be calculated from Eq. \ref{Bogoliubovcoefficient}. We use:
\begin{align}
    f_k(\tau \rightarrow -\infty) &= \frac{e^{-ik\tau}}{\sqrt{2k}}~~~~ {\rm Minkowski~mode ~function}\\
    f_k(\tau) &= \frac{e^{-ik\tau}}{\sqrt{2k}}\bigg(1-\frac{i}{k\tau}\bigg)~~~~ {\rm de~Sitter~ mode~ function}.
\end{align}
The time-dependent expressions for the Bogoliubov coefficients are:
\begin{align}
    \alpha(\tau)= 1-\frac{1}{2k^2\tau^2}-\frac{i}{k\tau}, ~~~~~ \beta(\tau)= \frac{e^{-2ik\tau}}{2k^2\tau^2}.
\end{align}
Introducing $y=-k\tau>0$, the Bogoliubov coefficient can be re-expressed as: 
\begin{align}
    \alpha(y)= 1-\frac{1}{2y^2}+\frac{i}{y}, ~~~ \beta(y)= \frac{e^{2iy}}{2y^2}.
\end{align}

\begin{figure}[h!]
    \centering
    \includegraphics[width=10cm]{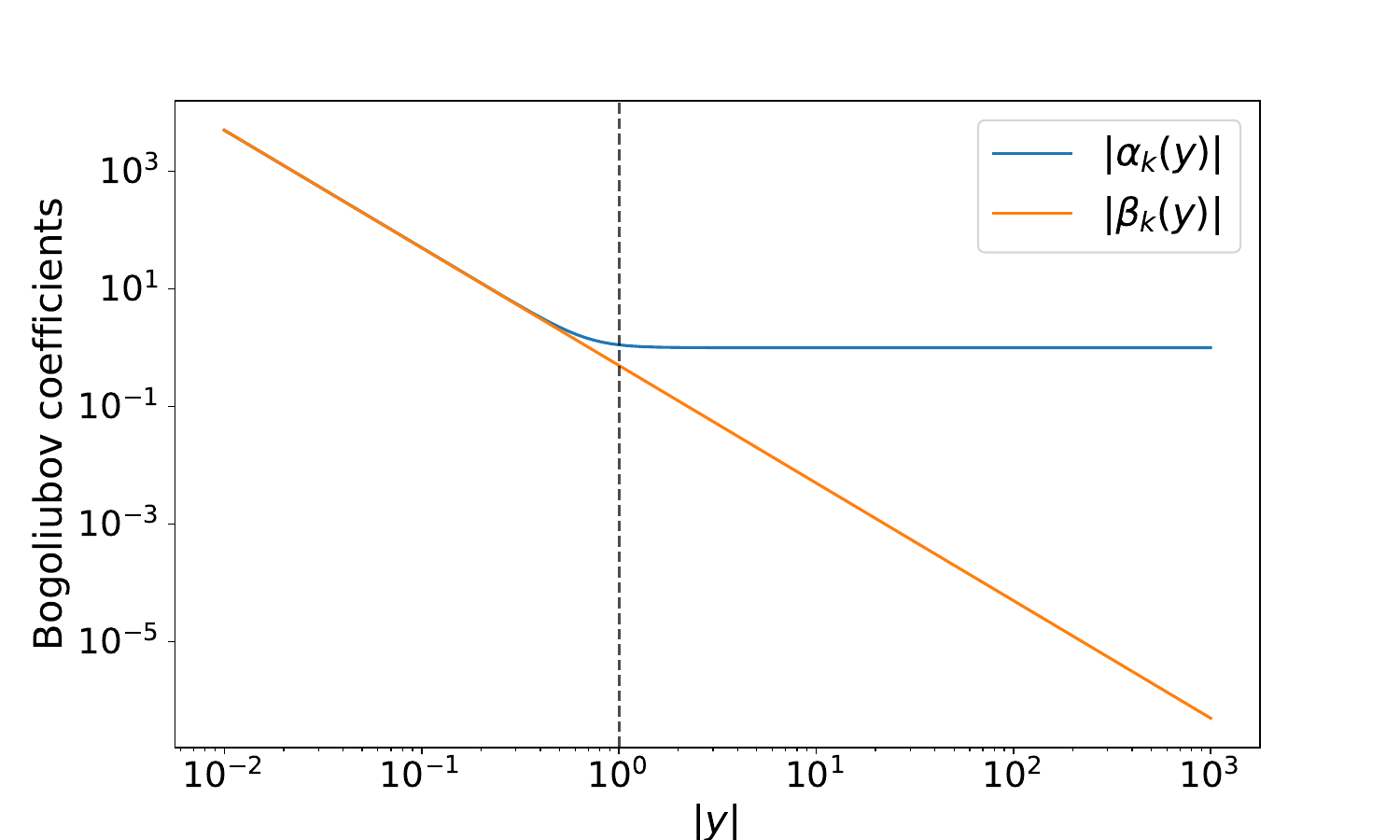}
    \caption{Log-Log plot of the absolute values of the Bogoliubov coefficients as a function of $|k\tau|$. For large values of $|k\tau|$, which corresponds to the early times, $|\beta_k| \rightarrow 0$ and $|\alpha_k| \rightarrow 1$, which is expected as we start from the Minkowski vacuum. The vertical dashed line separates the super-Hubble and the sub-Hubble regions. }
    \label{fig:enter-label}
\end{figure}

The following expressions will also be useful for the purpose of our analysis:
\begin{align}
    |\alpha_k| &= \sqrt{1+\frac{1}{4y^4}}, \\
    |\beta_k| &= \frac{1}{2y^2}, \\
    \tan {\rm arg}(\alpha_k) &= \frac{2y}{2y^2-1},\\
    \tan{\rm arg}(\beta_k) &= 2y, \\
    \tan{\rm arg}(\alpha_k \beta_k) &= \frac{2y \cos(2y)+(2y^2-1)\sin(2y)}{(2y^2-1)\cos(2y)-2y\sin(2y)},
\end{align}
from where:
\begin{align}
    r_k(\tau) &= {\rm arcsinh}|\beta_k(\tau)| = {\rm arcsinh} \left(\frac{1}{2y^2} \right),\\
    \theta_k(\tau) &= -{\rm arg} (\alpha_k(\tau)) = - \text{arctan}\left(\frac{2y}{2y^2-1}\right), \\
    \phi_k(\tau) &= {\rm arg}(\alpha_k(\tau)\beta_k(\tau))=\text{arctan}\left[\frac{2y \cos(2y)+(2y^2-1)\sin(2y)}{(2y^2-1)\cos(2y)-2y\sin(2y)} \right]. \label{alphabetatau}
\end{align}

\begin{figure}[h!]
    \centering
    \includegraphics[width=10cm]{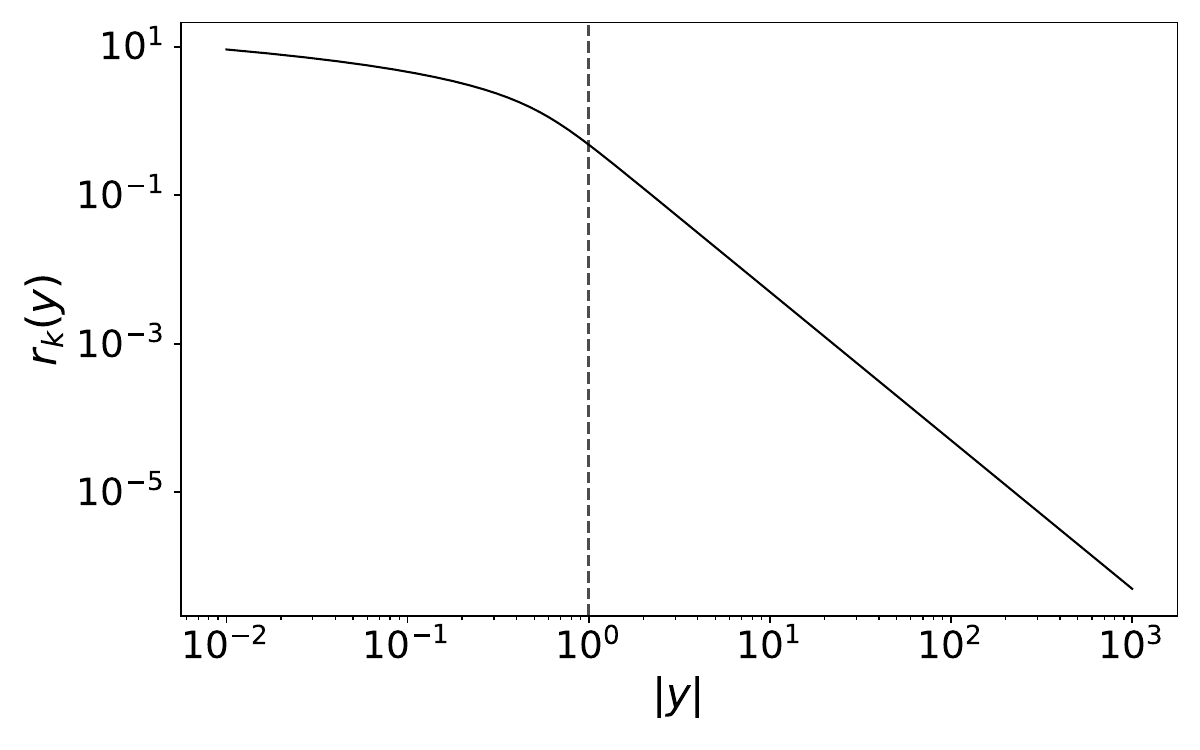}
    \caption{Log Log plot of the variation of the squeezing parameter as a function of $|k\tau|$. The vertical dashed line separates the super-Hubble and the sub-Hubble regions.}
    \label{squeezedparameter}
\end{figure}

In Fig. \ref{squeezedparameter}, the time-dependence of the 
squeezing amplitude is shown, exhibiting growth towards the 
super-Hubble scales. 

At $\tau \rightarrow -\infty$, i.e., $y \rightarrow \infty$, $\beta(y) \rightarrow 0$,  
so that:
\begin{align}
    r_k(y\rightarrow \infty)= 0, ~~~~ \theta_k(y \rightarrow \infty) = 2 m \pi.
\end{align}
Eq. \ref{alphabetatau}, shows that $\phi_k(y \rightarrow \infty)$ is undetermined in this limit. The quantity $\phi(\infty)$ represents the starting point in the unitary group manifold, which essentially gives us the freedom to choose a value for $\phi(\infty)$. 
Having all the essential quantities in hand, the complexity of the evolution of a scalar field mode in de Sitter background is given by:
\begin{align}
\label{complexitydesittermatrix}
    C[U_{\rm evolution}]= |v_3(y)|\sqrt{1+\tanh^2(r_k(y))\csc^2(2\theta_k(y)-\phi_k(y)+\phi_k(\infty)-2m\pi)}.
\end{align}

\begin{figure}[h!]
    \centering
    \includegraphics[width=10cm]{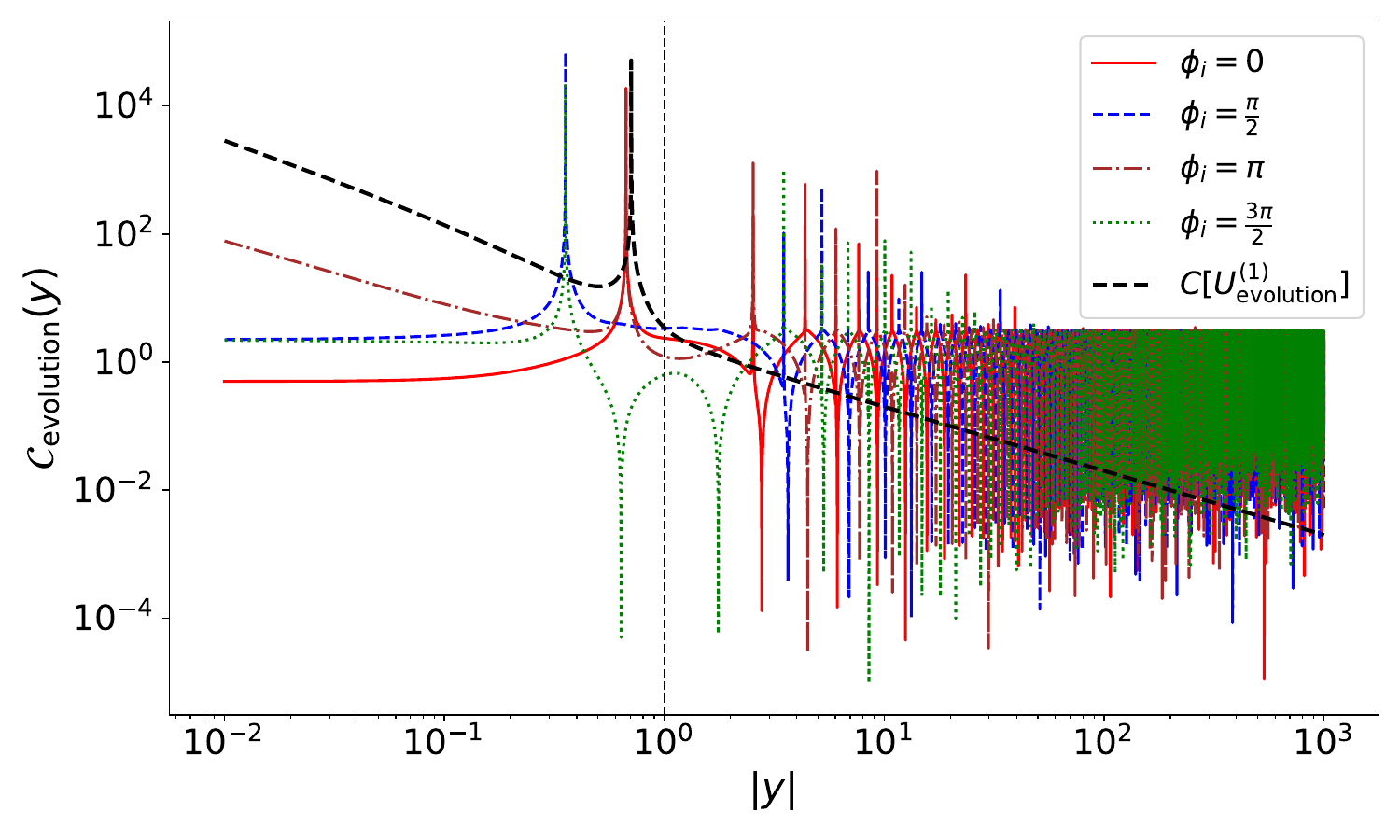}
    \caption{Log-Log plot of the geometric complexity of evolution of a scalar field in de Sitter spacetime for different values of the initial angle $\phi(\infty)$. The black dashed curve shows the complexity result derived using the approximations used while dealing with the path ordered and the operator expansion. The dashed vertical line separates the sub-Hubble and the super-Hubble limits.}
    \label{fig:enter-label}
\end{figure}

Let us study the $y\rightarrow \infty$ limit, which is the early time limit, and 
see what the value of the complexity is. In this limit, we have:
\begin{align}
    r_k(y\rightarrow \infty)=0,  ~~~~ \theta_k(y\rightarrow \infty) = 2 m \pi, ~~~~ \phi_k(y\rightarrow \infty) = \phi(\infty)~(\rm some~ unspecified ~quantity).
\end{align}
Substituting in Eq. \ref{complexitydesittermatrix}, we get:
\begin{align}
    C[U_{\rm evolution}]= |v_3(y\rightarrow \infty)| \rightarrow 0 .
\end{align}

The complexity formula written in \ref{complexitydesittermatrix} derived using the matrix representation of the generators shows striking differences with the formula derived following a representation-independent approach in Eq. \ref{upperbound} and \ref{improvedupperbound}. In the upper bound analysis, complexity was found to be independent of the squeezing angle $\phi$ and depended only on the squeezing parameter $r$ and the rotation parameter $\theta$. However, the formula written in \ref{complexitydesittermatrix} shows that the complexity is not independent of the squeezing angle $\phi$. Furthermore, it also shows that the complexity varies as $\tanh(r_k)$ with respect to the squeezing parameter $r_k$.

The quantity $\phi_k$ is a matter of choice. One way of taking care of it is to average over all possible values of $\phi_k(\infty)$:
\begin{align}
    \langle C_{\rm evolution} \rangle &= \frac{1}{2\pi}\int_{\phi(\infty) \in [0,2\pi)}C_{\rm evolution}(\phi(\infty))d\phi(\infty).
\end{align}

This procedure is not unique, and other consistent prescriptions for handling the undetermined quantity may exist.
 
\section{Application 2: An asymptotically static universe}
\label{sec6}

Let us consider a universe with the following 
scale factor dependence on conformal time $\tau$:
\begin{align}
    a^2(\tau)= \frac{1}{2}(a_f^2+
    a_i^2+(a_f^2-a_i^2)\tanh(\rho \tau)) .
\end{align}

The above scale factor describes a universe that evolves 
from an initial scale factor $a_i$ to $a_f$. This model was 
extensively reviewed in \cite{Birrell:1982ix}.

\begin{figure}[h!]
    \centering
    \includegraphics[scale=0.36]{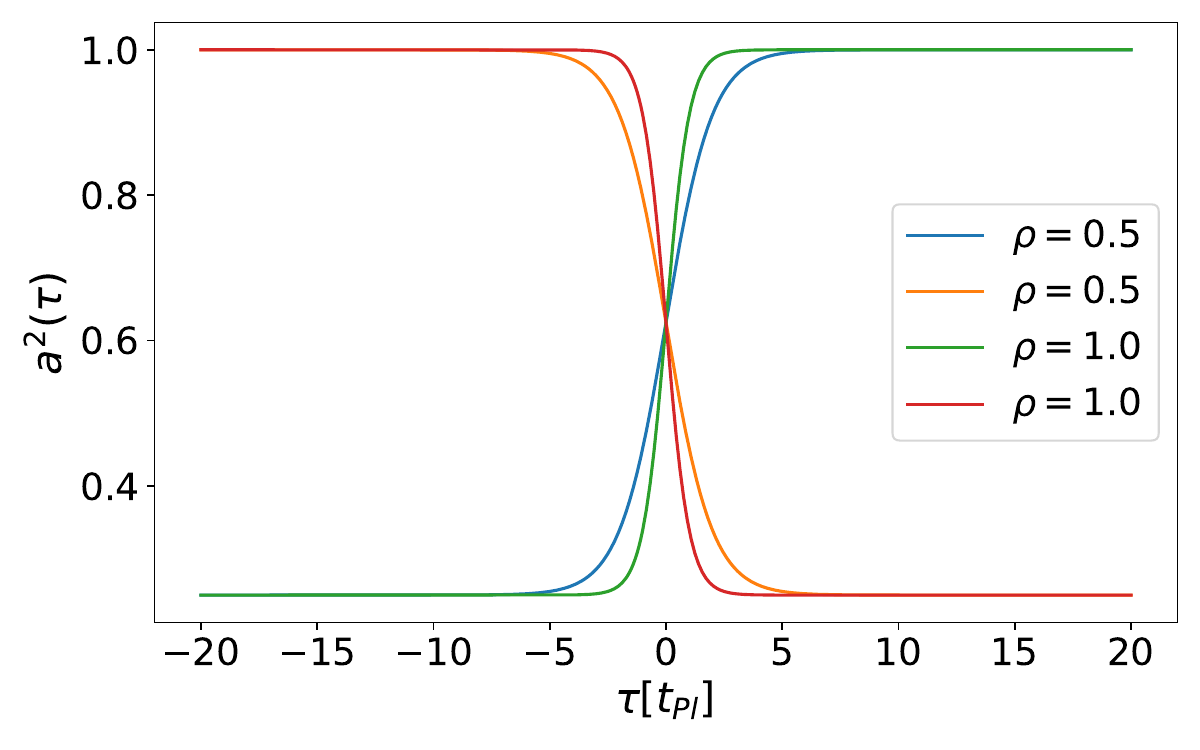}
    \includegraphics[scale=0.36]{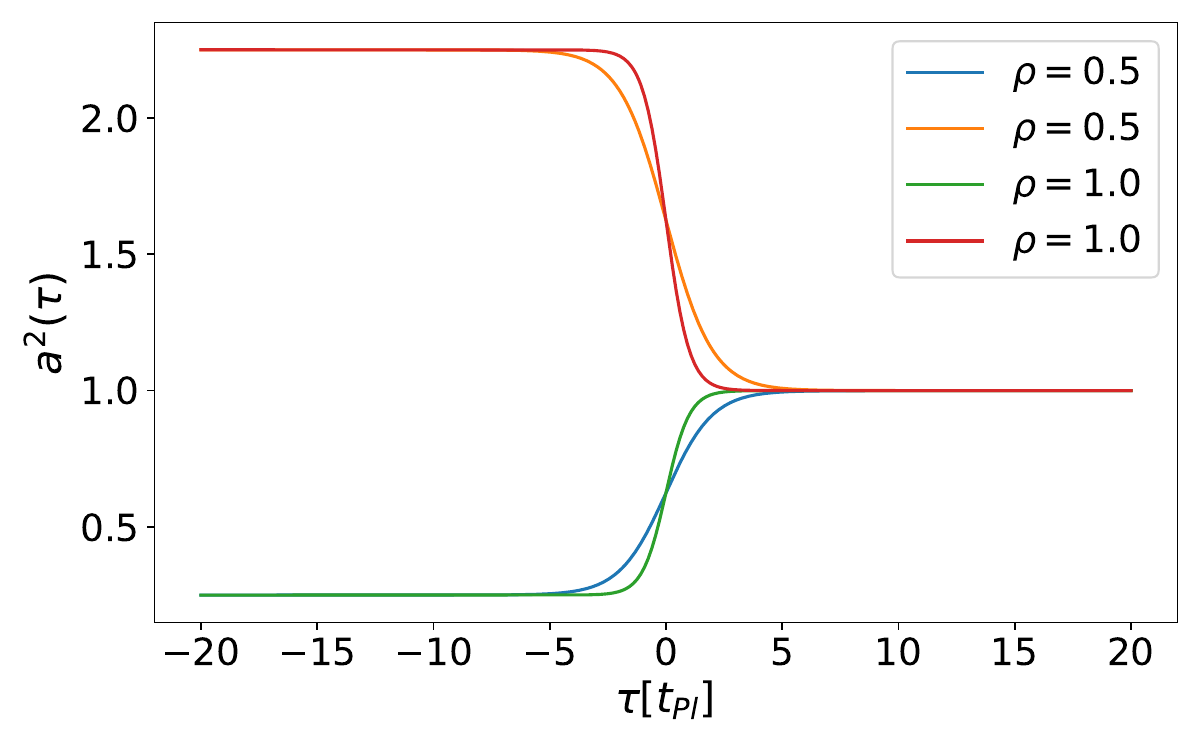}
    \caption{Illustration of the scale factor for an asymptotically bounded expanding and contracting universe. The spacetime is flat in the past with $a=a_i$. The universe then expands or contracts for a finite interval of the conformal time $\tau$. In the future, the spacetime is again flat with $a=a_f$. In the first figure, we consider a situation where, in the expanding case, the universe expands from a scale factor $a_i$ to $a_f$. Then, in the contracting case, the values of $a_i$ and $a_f$ are interchanged. In the second case,we consider a situation where the final scale factor $a_f$ is fixed at . For the expanding case, we consider $a_i=0.5$, and for the contracting case, we take $a_i=1.5$.  }
    \label{figscalefactor}
\end{figure}

The mass profile of the scalar field in this spacetime is given by:
\begin{align}
    M^2(\tau)= m^2 a^2 = \frac{m^2}{2}(a_f^2+
    a_i^2+(a_f^2-a_i^2)\tanh(\rho \tau)),
\end{align}
where $\rho$ denotes the time scale in which the spacetime interpolates 
from an initial scale factor $a_i$ at $\tau \rightarrow -\infty$ to the 
final value $a_{f}$ at $\tau \rightarrow \infty$. Precisely, it represents the inverse of the expansion time, i.e., a smaller value of $\rho$ indicates 
a greater expansion time.

The masses in the asymptotic regions can be written as:
    \begin{align}
M^2= 
    \begin{cases}
     M_{\rm in}^2 = m^2 a_i^2, ~~~~~~ \tau \rightarrow -\infty \\
     M_{\rm out}^2 = m^2 a_f^2, ~~~~~~ \tau \rightarrow \infty
\end{cases}
\end{align}

The ``in'' and ``out'' mode functions can be written as \cite{Birrell:1982ix, Ford:2021syk}: 
\begin{align}
    f_k^{\rm in} &= \frac{1}{\sqrt{2\omega_{\rm in}}}\exp\bigg(-i\omega_{+}\tau-i\frac{\omega_{-}}{\rho} \ln[{2\cosh(\rho \tau)}]\bigg) \nonumber \\ 
    &\times {}_2F_1\bigg(1+ i\frac{\omega_{-}}{\rho}, i\frac{\omega_{-}}{\rho};1-\frac{i\omega_{\rm in}}{\rho}; \frac{1}{2}[1+\tanh(\rho \tau)]\bigg), \\
    f_{k}^{\rm out} &= \frac{1}{\sqrt{2\omega_{\rm out}}}\exp\bigg(-i\omega_{+}\tau-i\frac{\omega_{-}}{\rho} \ln[{2\cosh(\rho \tau)}]\bigg) \nonumber \\
    & \times {}_2F_1\bigg(1+ i\frac{\omega_{-}}{\rho}, i\frac{\omega_{-}}{\rho};1+\frac{i\omega_{\rm out}}{\rho}; \frac{1}{2}[1-\tanh(\rho \tau)]\bigg),
\end{align}
where ${}_2 F_1$ denotes the (Gauss) hypergeometric function and  
$\omega_{\pm}= \frac{\omega_{\rm out}\pm \omega_{\rm in}}{2}$. 
Here, $\omega_{\rm in}$ and $\omega_{\rm out}$ are given by: 
\begin{align}
    \omega_{\rm in} = \sqrt{k^2+a_i^2 m^2}, ~~~~~ \omega_{\rm out} = \sqrt{k^2+ a_f^2 m^2}. 
\end{align}

In this context, the mode functions $f_k^{\rm in}(\tau)$ and $f_k^{\rm out}(\tau)$
will be referred to as ``in" modes and ``out" modes. In the asymptotic past 
($\tau \rightarrow -\infty$) and the asymptotic future ($\tau \rightarrow \infty$),
the modes are $\frac{1}{\sqrt{2\omega_{\rm in}}}e^{-i\omega_{\rm in}\tau}$ and $\frac{1}{\sqrt{2\omega_{\rm out}}}e^{-i\omega_{\rm out}\tau}$ respectively. 

The Bogoliubov coefficients relating the two solutions can be written as \cite{Ford:2021syk}: 
\begin{align}
    \alpha_k &= \sqrt{\frac{\omega_{\rm out}}{\omega_{\rm in}}}\frac{\Gamma\left(1-i \frac{\omega_{\rm in}}{\rho}\right)\Gamma\left(-i \frac{\omega_{\rm out}}{\rho}\right)}{\Gamma\left(1-i \frac{\omega_{+}}{\rho}\right)\Gamma\left(-i \frac{\omega_{+}}{\rho}\right)}, \\
    \beta_k &= \sqrt{\frac{\omega_{\rm out}}{\omega_{\rm in}}}\frac{\Gamma\left(1-i \frac{\omega_{\rm in}}{\rho}\right)\Gamma\left(i \frac{\omega_{\rm out}}{\rho}\right)}{\Gamma\left(1+i \frac{\omega_{-}}{\rho}\right)\Gamma\left(i \frac{\omega_{-}}{\rho}\right)}.
\end{align}

In the following, we will study the complexity of the ``out'' vacuum with respect to the ``in'' vacuum. This tells us how ``complex" the state in the asymptotic future would look to an observer in the asymptotic past and vice versa. The ``out'' vacuum state is related to the ``in'' vacuum state by the squeezed operator:
\begin{align}
    \ket{0_{\rm out}}= \hat{S}_2(r,\phi) \hat{R}_2(\theta)\ket{0_{\rm in}}.
\end{align}
Considering the action of the rotation operator on the vacuum state, we can write: 
\begin{align}
    \hat{S}_2(r,\phi)\hat{R}_2(\theta)\ket{0_{\rm in}}= e^{i\theta}\hat{S}_2(r,\phi)\ket{0_{\rm in}}.
\end{align}
The ``out'' vacuum in general can have a phase difference with respect to 
the ``in'' vacuum, which contributes to the complexity measure. However, 
to simplify the situation, we restrict our analysis in this article by 
considering $\theta=0$ and take into account only the contribution coming 
due the squeezing effect.   

Therefore, the unitary operator whose complexity would be of interest to us in this case is $S_2(r,\phi)$, whose complexity is given by:
\begin{align}
    C[\ket{0_{\rm out}}\rightarrow \ket{0_{\rm in}}]= |\tanh(r_k)|.
\end{align}

We have:
\begin{align}
    r_k &= {\rm arsinh|\beta_k|} \implies \sinh(r_k)= |\beta_k| \nonumber \\
    & \cosh^2(r_k)= 1+|\beta_k|^2 = |\alpha_k|^2 \nonumber\\
    & \tanh(r_k)= \frac{|\beta_k|}{|\alpha_k|}= \left|\frac{\beta_k}{\alpha_k}\right|.
\end{align}

Therefore, the complexity of $\hat{U}_{\rm target}$ can be written as:
\begin{align}
    C[\ket{0_{\rm out}}\rightarrow \ket{0_{\rm in}}]= \left|\frac{\beta_k}{\alpha_k}\right|= \left|\frac{\Gamma(i \frac{\omega_{\rm out}}{\rho})\Gamma(1-i \frac{\omega_{+}}{\rho})\Gamma(-i \frac{\omega_{+}}{\rho})}{\Gamma(-i \frac{\omega_{\rm out}}{\rho})\Gamma(1+i \frac{\omega_{-}}{\rho})\Gamma(i \frac{\omega_{-}}{\rho})}\right|.
\end{align}

It is instructive to simplify the above equation. Using the following properties 
of the $\Gamma$ function:
\begin{align}
    |\Gamma(i b)|^2 &= \frac{\pi}{b \sinh(\pi b)},\\
    |\Gamma(1+ib)|^2 &= \frac{\pi b}{\sinh(\pi b)},
\end{align}
the complexity $C[\ket{0_{\rm out}}\rightarrow \ket{0_{\rm in}}]$ can be simplified to:
\begin{align}
\label{complexityformula}
    C[\ket{0_{\rm out}}\rightarrow \ket{0_{\rm in}}] &= \frac{\sinh \left(\frac{\pi \omega_{-}}{\rho}\right)}{\sinh\left(\frac{\pi \omega_{+}}{\rho}\right)}.
\end{align}

Depending on whether the universe is expanding or contracting, the sign of $\omega_{-}$ is positive or negative respectively, which can be understood as follows:
\begin{align}
    \omega_{-}= \frac{\omega_{\rm out}-\omega_{\rm in}}{2}= \frac{\sqrt{k^2+a_f^2m^2}-\sqrt{k^2+a_i^2m^2}}{2},
\end{align}
so that:
\begin{align}
    \omega_{-}=\begin{cases}
        +ve~~~    {\rm when }~ a_f>a_i ~~~ {\rm expanding ~ universe}\\
        -ve~~~    {\rm when }~ a_f<a_i ~~~ {\rm contracting ~ universe}\\
    \end{cases}~~.
\end{align}
To accurately capture this, the correct way to express the complexity formula written in \ref{complexityformula} is given by:
\begin{align}
    C[\ket{0_{\rm out}}\rightarrow \ket{0_{\rm in}}]= \frac{\sinh\left(\frac{\pi |\omega_{-}|}{2\rho}\right)}{\sinh\left(\frac{\pi \omega_{+}}{2\rho}\right)} = \frac{\sinh\left(\frac{\pi (|\omega_{\rm out}-\omega_{\rm in}|)}{2\rho}\right)}{\sinh \left(\frac{\pi (\omega_{\rm out}+\omega_{\rm in})}{2\rho}\right)}.
\end{align}
To be more specific, let us write it down in terms of $a_i$ and $a_f$.
\begin{align}
    C[\ket{0_{\rm out}}\rightarrow \ket{0_{\rm in}}]= \frac{\sinh\bigg(\frac{\pi (|\sqrt{k^2+a_f^2m^2}-\sqrt{k^2+a_i^2m^2}|)}{2\rho}\bigg)}{\sinh\bigg(\frac{\pi (\sqrt{k^2+a_f^2m^2}+\sqrt{k^2+a_i^2m^2})}{2\rho}\bigg)}.
\end{align}
We will now study the limiting cases:
\begin{itemize}
    \item When $\rho \rightarrow 0$ i.e for a universe which is undergoing change infinitely slowly, we have:
\end{itemize}
\begin{align}
    C[\ket{0_{\rm out}}\rightarrow \ket{0_{\rm in}}] &\approx e^{\frac{\pi |\omega_{-}|}{\rho}}e^{-\frac{\pi \omega_{+}}{\rho}}= e^{\frac{\pi(|\omega_{-}|-\omega_{+})}{\rho}}.\\
\end{align}
Therefore, we have:
\begin{align}
    C[\ket{0_{\rm out}}\rightarrow \ket{0_{\rm in}}] \approx \begin{cases}
        e^{-\frac{\pi \omega_{in}}{\rho}}~~~~ {\rm expanding~universe} \\
        e^{-\frac{\pi \omega_{out}}{\rho}}~~~~ {\rm contracting~universe}
    \end{cases}.
\end{align}
Similarly, for large values of $\rho$ i.e for a universe undergoing sudden transition, we have:
\begin{align}
    C[\ket{0_{\rm out}}\rightarrow \ket{0_{\rm in}}] \approx \frac{|\omega_{-}|}{\omega_{+}}.
\end{align}
Therefore, we have:
\begin{align}
    C[\ket{0_{\rm out}}\rightarrow \ket{0_{\rm in}}] \approx \begin{cases}
        \frac{\omega_{\rm out}-\omega_{\rm in}}{\omega_{\rm out}+\omega_{\rm in}}~~~ {\rm expanding ~universe} \\
        \frac{\omega_{\rm in}-\omega_{\rm out}}{\omega_{\rm out}+\omega_{\rm in}}~~~ {\rm contracting~ universe}
    \end{cases}.
\end{align}

\begin{figure}[h!]
    \centering
    \includegraphics[width=10cm]{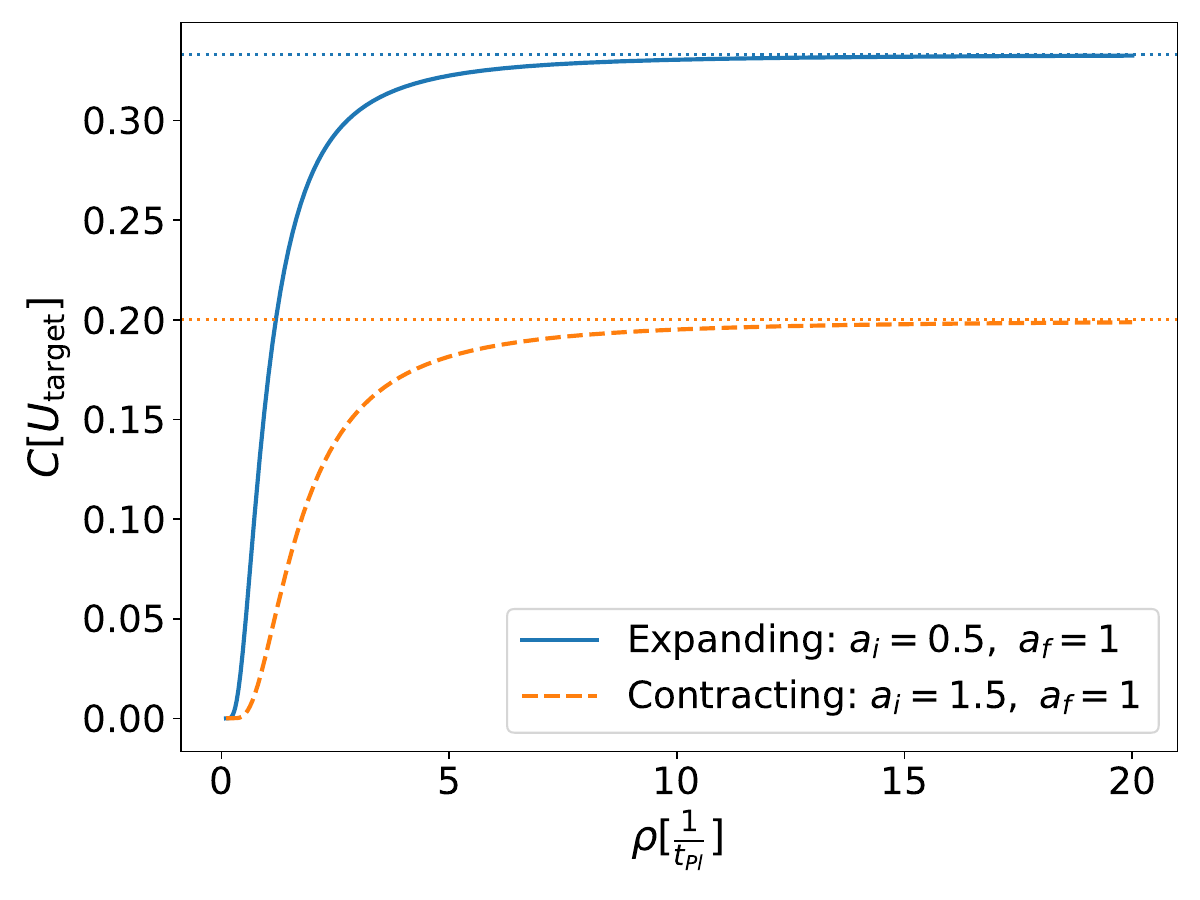}
    \caption{Complexity of the out vacuum with respect to the in vacuum. The figure illustrates the scenario in which the final scale factor $a_f$ is fixed and $a_i$ is varied for the expanding and contracting cases.}
    \label{complexityplotaiafinterchanged}
\end{figure}
  
From the perspective of an ``in" observer, the $\ket{0}_{\rm out}$ is a particle state, and the number of particles created in a given mode is given by:
\begin{align}
    n_k= |\beta_k|^2.
\end{align}
This allows us to give an interpretation of complexity in terms of the number of particles as follows:
\begin{align}
    C[\ket{0_{\rm out}}\rightarrow \ket{0_{\rm in}}]= 
    \begin{cases}
        2 {\rm arsinh}(\sqrt{n_k}) ~~ \text{Upper bound result} \\
        \frac{\sqrt{n_k}}{\sqrt{1+n_k}}~~~~~~~~~~~~ \text{Exact complexity}
    \end{cases} ~.
\end{align}

\begin{figure}[h!]
    \centering
    \includegraphics[width=10cm]{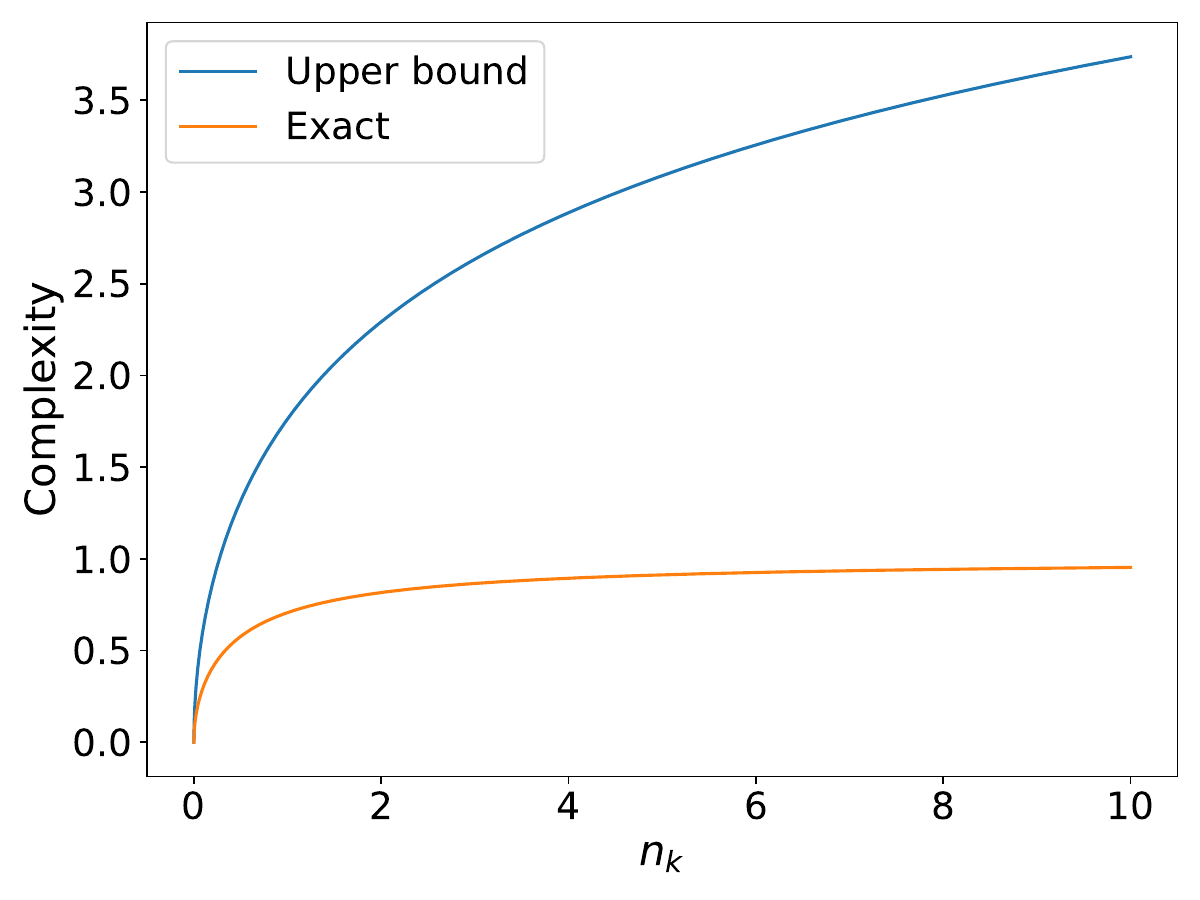}
    \caption{Variation of complexity as a function of the number of particles produced.  The upper bound is based on the formula 
    derived in Ref. \cite{Chowdhury:2024ntx}.}
    \label{Bernardduncanplotcompvsn}
\end{figure}

\section{Summary and discussions}
\label{sec7}

In the present work, we derive an exact expression for the complexity of cosmological 
perturbations. We adopt the geometric approach pioneered by Nielsen and collaborators, in
which the complexity of a unitary operator is defined as the geodesic distance between the
identity operator and the target unitary in an appropriate unitary group manifold. This 
group manifold is determined by the generators of the Lie algebra relevant to the unitary
operator under consideration.

The unitary operator appropriate for studying the evolution of cosmological perturbations 
is given by the product of a two-mode squeezing operator and a rotation operator. The 
fundamental operators used to construct this unitary were shown to satisfy the $\mathfrak{su}
(1,1)$ Lie algebra. Employing a well-known finite-dimensional matrix representation of the
$SU(1,1)$ group generators, we derive an explicit formula for the geometric complexity. 
This finite-dimensional representation enables us to go beyond the upper-bound approximations
previously employed in Refs.~\cite{Chowdhury:2023iwg,Chowdhury:2024ntx}.

Within the geometric quantum complexity framework, the trajectory connecting the identity 
and the target unitary operator on the group manifold is represented by a path-ordered
exponential. A standard method for evaluating such path-ordered exponentials is to express
them in terms of a Dyson series. In principle, an exact evaluation requires the inclusion of
all terms in the Dyson series; however, this becomes technically challenging when the Lie
algebra of the generators does not possess a center, i.e., when no element commutes with all
others. In such cases, the Dyson series contains infinitely many terms.

In Ref.~\cite{Chowdhury:2024ntx}, the path-ordered exponential was evaluated by 
retaining only the leading-order term in the Dyson series. While this approximation 
yields valuable qualitative insights, the resulting expression can be interpreted only 
as an upper bound on the complexity. To obtain the exact value of the complexity, it is
therefore necessary to incorporate the contributions from all higher-order terms in the 
Dyson series. One systematic way to achieve this is to work with a finite-dimensional
representation of the Lie algebra generators, which allows us to directly solve the 
time-dependent Schr\"odinger equation satisfied by the path-ordered exponential.

By considering a finite-dimensional matrix representation of the generators and 
a generic group element, we derived the complexity of a general unitary operator
belonging to the $SU(1,1)$ group. We find that the resulting complexity is determined
up to an a priori unspecified contribution, which is associated with the phase of the 
$SU(1,1)$ group element. Although one might be tempted to neglect the contribution 
of such phase factors in defining the complexity measure, our analysis shows that 
they contribute in a nontrivial manner. From a geometric perspective, these phase 
factors encode the orientation of the group element on the group manifold and 
therefore constitute an essential part of the complexity measure.

Having obtained the general expression for the complexity, we subsequently apply 
this formalism to compute the complexity of physically relevant unitary operators 
constructed from $\mathfrak{su}(1,1)$ Lie algebra generators.

We begin by analyzing the complexity of the two-mode rotation and squeezing 
operators separately, and subsequently study their product, which constitutes 
the time-evolution operator for scalar field modes on dynamical backgrounds. 
The exact complexity of the squeezing operator depends on both the squeezing a
mplitude $r_k$ and the squeezing angle $\phi_k$. In contrast, the upper-bound r
esult derived in Ref.~\cite{Chowdhury:2024ntx} fails to capture the dependence 
on $\phi_k$. This demonstrates that the squeezing angle plays a role in the 
complexity measure that is as significant as that of the squeezing amplitude $r_k$.

This observation is particularly important in the context of quantum dynamics 
and becomes more transparent when the combined effect of the squeezing and 
rotation operators is taken into account in the study of scalar field evolution 
on cosmological spacetimes. Another notable deviation of the exact complexity 
from the upper-bound approximation appears in its dependence on the squeezing 
amplitude. Specifically, the exact complexity scales as $C \sim \tanh(r_k)$, 
rather than linearly with $r_k$ as predicted by the upper-bound result.

Furthermore, our analysis of the product of the squeezing and rotation operators 
reveals that the complexity depends on both the squeezing and rotation parameters. 
A crucial additional observation is that the complexity is sensitive to the initial 
value of the squeezing angle, indicating that phase information contributes nontrivially 
to the complexity measure. Finally, the explicit dependence of the complexity of 
evolution on the rotation parameter $\theta_k(t)$ suggests that it cannot be neglected, 
in contrast to earlier state-based approaches, where this contribution was ignored
\cite{Bhattacharyya:2020kgu,Bhattacharyya:2020rpy,Lehners:2020pem}.

We revisit the complexity of the time-evolution operator for both the harmonic 
oscillator and the inverted harmonic oscillator. For the harmonic oscillator, 
we recover the expected oscillatory behavior of the complexity. In contrast, 
the inverted harmonic oscillator exhibits a marked deviation from the upper-bound 
result. In particular, the upper bound derived in Ref.~\cite{Chowdhury:2023iwg}
predicts linear growth of the complexity at all times, whereas the exact expression 
reveals a saturation of complexity at late times.

We then apply the general formalism to scalar fields propagating on cosmological 
spacetimes. As a first example, we consider a massless test scalar field on a de 
Sitter background. Our analysis demonstrates that the complexity is highly sensitive 
to the initial value of the squeezing angle. This sensitivity becomes especially 
pronounced in the super-Hubble regime. Interestingly, for initial squeezing 
angles $\phi_k \sim n\pi$ ($n>0$), the complexity exhibits a growing behavior 
in the super-Hubble regime, in contrast to other values of $\phi_k$ for which 
the complexity saturates. In the sub-Hubble regime, on the other hand, the complexity 
displays an oscillatory behavior. This behavior can be directly understood from the 
frequency profile of the field modes given in Eq.~\ref{desitterfrequency}. For 
$k\tau > \sqrt{2}$, the field modes behave as standard harmonic oscillators, 
leading to oscillatory complexity. However, when $k\tau < \sqrt{2}$, the effective 
frequency satisfies $\omega_{\text{dS}}^2 < 0$, corresponding to an imaginary frequency,
and the field modes resemble inverted harmonic oscillators, resulting in a similar 
complexity pattern.

This analysis further reveals that, in the sub-Hubble regime, the evolution of complexity is dominated by the rotation operator. In contrast, in the super-Hubble regime, the squeezing operator plays the dominant role.

As a second application, we study an asymptotically static universe, characterized 
by constant scale factors in both the asymptotic past and the asymptotic future. 
This model is particularly advantageous from multiple perspectives. First, since 
the spacetime is static in both asymptotic regions, well-defined stationary states 
of the quantum fields exist, allowing for a clear definition of particles. This makes 
the model especially useful for investigating cosmological particle production. 
Second, it enables the study of two distinct scenarios—an expanding or a contracting 
universe—depending on whether the final scale factor is larger or smaller than the 
initial one.

In this work, we restrict our analysis to the complexity of the vacuum state in the 
``out'' region relative to the ``in'' vacuum. We do not address the time evolution 
of the complexity of individual field modes, which is an interesting direction that 
we leave for future investigation. Our results indicate that the complexity depends 
on the difference between the initial and final scale factors, irrespective of whether 
the universe undergoes expansion or contraction. Another important observation is 
the sensitivity of complexity to the rate at which the universe evolves. In the limit 
of an infinitely slow evolution, the behavior of complexity differs qualitatively 
between expanding and contracting universes.
           
In conclusion, the analysis presented in this article addresses several key questions 
concerning the deviation of the true complexity from that obtained using the upper-bound
approximation. Moreover, it establishes geometric complexity as a sensitive probe 
capable of capturing intricate features of the underlying cosmological spacetime. 
This, in turn, opens new avenues for exploring the early Universe through the lens of 
quantum complexity, providing insight into the emergence and evolution of complexity in
cosmology.

\appendix

\section{Derivation of the exact solutions}

Using the identity given in Eq.~\eqref{Eqncothx}, Eq.~\eqref{dyds} can be 
recast in the form:
\begin{align}
\frac{dy}{ds} &= v_3 \pm \frac{1}{2v_3}(v_3^2+K^2) \nonumber \\
&=
v_3 \pm \frac{1}{2v_3}\left[v_3^2+\frac{1}{2}(v_1^2+v_2^2)+\frac{1}{2}(v_1^2-v_2^2)\cos (2y)
-v_1v_2 \sin (2y) \right].
\end{align}

By collecting all constant terms, the above equation can be written in the compact form:
\begin{align}
\frac{dy}{ds}= a+ b \cos(2y)+ c \sin(2y),
\end{align}
where the coefficients $a$, $b$, and $c$ are defined as
\begin{align}
a &= v_3 \pm \frac{1}{2 v_3}\bigg(v_3^2+\frac{1}{2}(v_1^2+v_2^2)\bigg), \\
b &= \pm \frac{1}{4 v_3}(v_1^2-v_2^2),\\
c &= \mp \frac{1}{2 v_3} v_1 v_2.
\end{align}

The differential equation can now be directly integrated, yielding:
\begin{align}
\int \frac{dy}{a+b\cos(2y)+c\sin(2y)}= \int ds.
\end{align}
Evaluating the integral leads to:
\begin{align}
\label{sintermsofy}
s= -\frac{\tanh ^{-1}\left(\frac{(a-b) \tan (y)+c}{\sqrt{b^2+c^2-a^2}}\right)}{\sqrt{b^2+c^2-a^2}}+d,
\end{align}
where $d$ is an integration constant.

For convenience, we introduce the notation:
\begin{align}
D := \sqrt{b^2+c^2-a^2}.
\end{align}
With this definition, Eq.~\eqref{sintermsofy} can be inverted to 
express $y$ explicitly as a function of $s$:
\begin{align}
\tan(y(s))= \frac{-c+ D \tanh(D(d-s))}{a-b},
\end{align}
which implies
\begin{align}
y(s)= \arctan\bigg(\frac{-c+ D \tanh(D(d-s))}{a-b}\bigg).
\end{align}

Recalling that $y(s)$ can be decomposed as:
\begin{align}
y(s)= 2s v_3+ \chi(s)+ \psi(s),
\end{align}
and using the solution for $\psi(s)$,
\begin{align}
\psi(s)= -v_3 s+ v,
\end{align}
we obtain an explicit expression for $\chi(s)$:
\begin{align}
\nonumber
\chi(s) &= \arctan\bigg(\frac{-c+ D \tanh(D(d-s))}{a-b}\bigg) - 2 sv_3 +v_3 s -v\\
&= \arctan\bigg(\frac{-c+ D \tanh(D(d-s))}{a-b}\bigg)- s v_3 -v.
\end{align}

Having determined $y(s)$, we now proceed to solve the equation governing $x(s)$,
\begin{align}
\frac{dx}{ds} &= \cos y (v_1 \tan y+ v_2) \nonumber\\
&= \pm \sqrt{\frac{1}{1+ \tan^2 y}}(v_1 \tan y+ v_2).
\end{align}

The solution to this equation can be expressed as:
\begin{align}
x(s)= C \pm \frac{\frac{{\rm arctanh}(\mathcal{M}) (v_2 (b-a)+c v_1-D v_1)}{\sqrt{a^2-2 a b+b^2+(c-D)^2}}-\frac{{\rm arctanh}(\mathcal{N}) (v_2 (b-a)+c v_1+D v_1)}{\sqrt{a^2-2 a b+b^2+(c+D)^2}}}{2 D},
\end{align}
where $C$ is an integration constant and the auxiliary functions $\mathcal{M}$ and $\mathcal{N}$ are defined as:
\begin{align}
\mathcal{M} &= \frac{a^2-2 a b+b^2+c^2-D (c-D) \tanh (D (d-s))-c D}{(a-b) \sqrt{a^2-2 a b+b^2+(c-D)^2} \sqrt{\frac{a^2-2 a b+b^2+c^2-2 c D \tanh (D (d-s))+D^2 \tanh ^2(D(d-s))}{(a-b)^2}}},\\
\mathcal{N} &= \frac{a^2-2 a b+b^2+c^2-D (c+D) \tanh (D (d-s))+c D}{(a-b) \sqrt{a^2-2 a b+b^2+(c+D)^2} \sqrt{\frac{a^2-2 a b+b^2+c^2-2 c D \tanh (D (d-s))+D^2 \tanh ^2(D (d-s))}{(a-b)^2}}}.
\end{align}

Finally, the function $\rho(s)$ can be obtained directly from $x(s)$ and is given by:
\begin{align}
\rho(s)= \frac{C}{2} \pm \frac{\frac{{\rm arctanh}(\mathcal{M}) (v_2 (b-a)+c v_1-D v_1)}{\sqrt{a^2-2 a b+b^2+(c-D)^2}}-\frac{{\rm arctanh}(\mathcal{N}) (v_2 (b-a)+c v_1+D v_1)}{\sqrt{a^2-2 a b+b^2+(c+D)^2}}}{4 D}.
\end{align}

\section{Derivation of the expression of the line element from the spinorial representation of $\mathfrak{su(1,1)}$}

The Lie group SU(1,1) is defined as the set of $2 \times 2$ matrices U of determinant 1 that satisfy the relation: 
\begin{align} U \epsilon U^{\dagger}= \epsilon, 
\end{align} 
where: 
\begin{align} 
\epsilon= \begin{pmatrix} 1 && 0 \\ 0 && -1 \end{pmatrix}. 
\end{align} 

Explicitly, the group element can be written as: 
\begin{align} 
U= \begin{pmatrix} \alpha && \beta \\ \bar{\beta} && \bar{\alpha} \end{pmatrix} 
, ~~~~ {\rm with} |\alpha|^2-|\beta|^2=1. 
\end{align} T
These properties actually allow us to use a parametrization of the 
group element such that: 
\begin{align} 
U= \begin{pmatrix} \cosh(\rho) e^{i\chi} && -\sinh(\rho) e^{i\psi} \\ -\sinh(\rho) e^{-i\psi} && \cosh(\rho) e^{-i\chi} \end{pmatrix}. 
\end{align} 

We can write down: 
\begin{align} 
idU U^{-1}= \mathcal{O}_I dx^I. 
\end{align} 

Since we have a matrix representation of the $O_I$'s, we can multiply 
the above equation by the inverse of $O_I$ from the right, which gives: 
\begin{align} 
idUU^{-1}\mathcal{O}_I^{-1}= \mathcal{O}_I\mathcal{O}_I^{-1}dx^I. 
\end{align} 

From this, we can write down: 
\begin{align} 
dx^I= \frac{1}{{\rm Tr}[\mathcal{O}_I\mathcal{O}_I^{-1}]}\bigg[{\rm Tr} (i dUU^{-1}\mathcal{O}_I^{-1})\bigg]. 
\end{align} 

Therefore, the line element with the metric $G_{IJ}$, 
can be written as: 
\begin{align} 
ds^2= G_{IJ}dx^I dx^J. 
\end{align} 

Using the above parametrization, we get 
\begin{align} 
dx^1 &= 2 \sin (\chi +\psi )d\rho + \sinh (2\rho) \cos (\chi +\psi )(d\psi-d\chi), \\ 
dx^2 &= 2\cos (\chi +\psi ) d\rho+ \sinh (2\rho) \sin (\chi +\psi )d\chi- \sinh (2\rho)\sin (\chi +\psi)d\psi, \\ 
dx^3 &= 2 \sinh ^2(\rho)d\psi-2\cosh ^2(\rho)d\chi. 
\end{align} 

With the choice $G_{IJ}=\delta_{IJ}$, the line element can be written as: \
\begin{align} 
\nonumber ds^2 &= 4 d\rho^2+2 d\chi^2 \cosh (2 \rho)+d\chi^2 \cosh (4 \rho)+d\chi^2-2 d\chi d\psi \cosh (4 \rho) \\ & ~~~~ +2 d\chi d\psi-2 d\psi^2 \cosh (2 \rho)+d\psi^2 \cosh (4 \rho)+d\psi^2. 
\end{align}

\section{Derivation of upper bound on complexity of the target unitary operator from a representation independent approach}
\label{appendixupperbound}

The operators $\mathcal{O}_i$'s satisfies the $\mathfrak{su(1,1)}$ lie algebra. The Euler-Arnold equations (with $G_{IJ}=\delta_{IJ}$) can be written as 
\begin{align}
    \frac{dV^1}{ds} &= -2 V^2V^3, \\
    \frac{dV^2}{ds} &= 2 V^1V^3, \\
    \frac{dV^3}{ds} &= 0,
\end{align}
which can be solved as 
\begin{align}
    V^1(s) &= v_1 \cos(2 s v_3)-v_2 \sin(2 s v_3),\\
    V^2(s) &= v_2 \cos(2 s v_3)+ v_1 \sin(2 s v_3), \\
    V^3(s) &= v_3.
\end{align}

The complexity of the target unitary operator in terms of the $v_i$'s 
can be written as: 
\begin{align}
C[U_{\rm target}]= \sqrt{v_1^2+v_2^2+v_3^2}.
\end{align}

We want to know the geodesic for fixed boundary conditions 
$U(s=0)= \mathbb{I}$ and $U(s)= U_{\rm target}$ to fix the $v_i$'s. 
The unitary along the geodesic path from the identity with a specific 
tangent vector $V(s)$ is given by the path-ordered exponential:
\begin{align}
    U(s)= \mathcal{P} \exp\bigg(-i\int_0^s V^I(s') \mathcal{O}_I ds'\bigg),
\end{align}
which is a solution to the equation: 
\begin{align}
\label{differentialU1}
    \frac{dU(s)}{ds}=-i V^I(s)\mathcal{O}_I U(s). 
\end{align}

We would like to solve it and see what $U(1)$ looks like as a function 
of $v_i$'s and then use the boundary condition $U(1)=U_{\rm target}$ in 
order to derive the $v_i$'s for a specified target unitary operator. 
However, solving $U(s), $ would require dealing with the path ordering, 
which is a notoriously difficult problem and is usually solved using an 
iterative approach, and the solution is usually expressed as 
\textit{Dyson series}.

Therefore, the path-ordered exponential can be written as: 
\begin{align}
    U(s)= \mathbb{I}-i \int_{0}^{s}V^I(s')\mathcal{O}_I ds' + (-i)^2 \int_{0}^{s} V^I(s') \mathcal{O}_I ds'\int_{0}^{s'}V^I(s'')\mathcal{O}_I ds''+.....
\end{align}

We will keep only the leading order term in the \textit{Dyson series}, which 
means that we will only be able to comment on the \textit{upper bound} of complexity rather than the precise value. 
Substituting the $V^I(s)$ obtained from the solution of the Euler-Arnold equation 
in the above equation, we can write: 
\begin{align}
    & \int_0^s V^I(s')\mathcal{O}_I ds' = \frac{\mathcal{O}_1 v_1 \sin (2 s v_3)}{2 v_3}-\frac{\mathcal{O}_1 v_2 \sin ^2(s v_3)}{v_3}+\frac{\mathcal{O}_2 v_1 \sin ^2(s v_3)}{v_3}+\frac{\mathcal{O}_2 v_2 \sin (2 s v_3)}{2 v_3}+\mathcal{O}_3 s v_3.
\end{align}

Keeping only up to the leading order term, $U(s)$ can be written as 
\begin{align}
\nonumber
    U(s) &\approx \exp\bigg(-i\bigg(\bigg\{\frac{v_1 \sin(2 s v_3)}{2 v_3}-\frac{v_2 \sin^2(s v_3)}{v_3}\bigg\}\mathcal{O}_1 \\ &~~~~~~~~~+\bigg\{\frac{v_2 \sin(2 s v_3)}{2 v_3}+\frac{v_1 \sin^2(s v_3)}{v_3}\bigg\}\mathcal{O}_2 +  s v_3 \mathcal{O}_3\bigg)\bigg).
\end{align}

To implement the boundary condition accurately, 
We can now implement the boundary condition
\begin{align}
    U(s=1)= S_2(r_k,\phi_k)= \exp\bigg[-2ir_k (\sin(\phi_k)\mathcal{O}_1+ \cos(\phi_k)\mathcal{O}_2)\bigg].
\end{align}

The condition leads to three equations:
\begin{align}
    v_3 &= 0, \\
    \frac{v_1 \sin(2 s v_3)}{2 v_3}-\frac{v_2 \sin^2(s v_3)}{v_3} & = 2r_k \sin(\phi_k), \\
    \frac{v_2 \sin(2 s v_3)}{2 v_3}+\frac{v_1 \sin^2(s v_3)}{v_3} &= 2r_k \cos(\phi_k), 
\end{align}
which gives the following values of $v_i's$:
\begin{align}
    v_1= 2 r_k \sin(\phi_k), ~~~ v_2= 2 r_k \cos(\phi_k).
\end{align}

\acknowledgments

SC would like to thank the Doctoral School of Exact and Natural Sciences 
of the Jagiellonian University for providing a fellowship during the course 
of the work. The research was conducted within the Quantum Cosmos Lab
(https://quantumcosmos.org/) at the Jagiellonian University. 

\bibliography{biblio}

\providecommand{\href}[2]{#2}\begingroup\raggedright\begin{thebibliography}{10}

\bibitem{Ryu:2006bv}
S.~Ryu and T.~Takayanagi, ``{Holographic derivation of entanglement entropy from AdS/CFT},'' \href{http://dx.doi.org/10.1103/PhysRevLett.96.181602}{{\em Phys. Rev. Lett.} {\bfseries 96} (2006) 181602}, \href{http://arxiv.org/abs/hep-th/0603001}{{\ttfamily arXiv:hep-th/0603001}}.

\bibitem{Hubeny:2007xt}
V.~E. Hubeny, M.~Rangamani, and T.~Takayanagi, ``{A Covariant holographic entanglement entropy proposal},'' \href{http://dx.doi.org/10.1088/1126-6708/2007/07/062}{{\em JHEP} {\bfseries 07} (2007) 062}, \href{http://arxiv.org/abs/0705.0016}{{\ttfamily arXiv:0705.0016 [hep-th]}}.

\bibitem{Susskind:2014moa}
L.~Susskind, ``{Entanglement is not enough},'' \href{http://dx.doi.org/10.1002/prop.201500095}{{\em Fortsch. Phys.} {\bfseries 64} (2016) 49--71}, \href{http://arxiv.org/abs/1411.0690}{{\ttfamily arXiv:1411.0690 [hep-th]}}.

\bibitem{Stanford:2014jda}
D.~Stanford and L.~Susskind, ``{Complexity and Shock Wave Geometries},'' \href{http://dx.doi.org/10.1103/PhysRevD.90.126007}{{\em Phys. Rev. D} {\bfseries 90} no.~12, (2014) 126007}, \href{http://arxiv.org/abs/1406.2678}{{\ttfamily arXiv:1406.2678 [hep-th]}}.

\bibitem{Brown:2015bva}
A.~R. Brown, D.~A. Roberts, L.~Susskind, B.~Swingle, and Y.~Zhao, ``{Holographic Complexity Equals Bulk Action?},'' \href{http://dx.doi.org/10.1103/PhysRevLett.116.191301}{{\em Phys. Rev. Lett.} {\bfseries 116} no.~19, (2016) 191301}, \href{http://arxiv.org/abs/1509.07876}{{\ttfamily arXiv:1509.07876 [hep-th]}}.

\bibitem{Carmi:2017jqz}
D.~Carmi, S.~Chapman, H.~Marrochio, R.~C. Myers, and S.~Sugishita, ``{On the Time Dependence of Holographic Complexity},'' \href{http://dx.doi.org/10.1007/JHEP11(2017)188}{{\em JHEP} {\bfseries 11} (2017) 188}, \href{http://arxiv.org/abs/1709.10184}{{\ttfamily arXiv:1709.10184 [hep-th]}}.

\bibitem{Swingle:2017zcd}
B.~Swingle and Y.~Wang, ``{Holographic Complexity of Einstein-Maxwell-Dilaton Gravity},'' \href{http://dx.doi.org/10.1007/JHEP09(2018)106}{{\em JHEP} {\bfseries 09} (2018) 106}, \href{http://arxiv.org/abs/1712.09826}{{\ttfamily arXiv:1712.09826 [hep-th]}}.

\bibitem{Fu:2018kcp}
Z.~Fu, A.~Maloney, D.~Marolf, H.~Maxfield, and Z.~Wang, ``{Holographic complexity is nonlocal},'' \href{http://dx.doi.org/10.1007/JHEP02(2018)072}{{\em JHEP} {\bfseries 02} (2018) 072}, \href{http://arxiv.org/abs/1801.01137}{{\ttfamily arXiv:1801.01137 [hep-th]}}.

\bibitem{Alishahiha:2018tep}
M.~Alishahiha, A.~Faraji~Astaneh, M.~R. Mohammadi~Mozaffar, and A.~Mollabashi, ``{Complexity Growth with Lifshitz Scaling and Hyperscaling Violation},'' \href{http://dx.doi.org/10.1007/JHEP07(2018)042}{{\em JHEP} {\bfseries 07} (2018) 042}, \href{http://arxiv.org/abs/1802.06740}{{\ttfamily arXiv:1802.06740 [hep-th]}}.

\bibitem{Chapman:2016hwi}
S.~Chapman, H.~Marrochio, and R.~C. Myers, ``{Complexity of Formation in Holography},'' \href{http://dx.doi.org/10.1007/JHEP01(2017)062}{{\em JHEP} {\bfseries 01} (2017) 062}, \href{http://arxiv.org/abs/1610.08063}{{\ttfamily arXiv:1610.08063 [hep-th]}}.

\bibitem{Yang:2017czx}
R.-Q. Yang, C.~Niu, C.-Y. Zhang, and K.-Y. Kim, ``{Comparison of holographic and field theoretic complexities for time dependent thermofield double states},'' \href{http://dx.doi.org/10.1007/JHEP02(2018)082}{{\em JHEP} {\bfseries 02} (2018) 082}, \href{http://arxiv.org/abs/1710.00600}{{\ttfamily arXiv:1710.00600 [hep-th]}}.

\bibitem{Couch:2017yil}
J.~Couch, S.~Eccles, W.~Fischler, and M.-L. Xiao, ``{Holographic complexity and noncommutative gauge theory},'' \href{http://dx.doi.org/10.1007/JHEP03(2018)108}{{\em JHEP} {\bfseries 03} (2018) 108}, \href{http://arxiv.org/abs/1710.07833}{{\ttfamily arXiv:1710.07833 [hep-th]}}.

\bibitem{Nielsen_2006}
M.~A. Nielsen, M.~R. Dowling, M.~Gu, and A.~C. Doherty, ``Quantum computation as geometry,'' \href{http://dx.doi.org/10.1126/science.1121541}{{\em Science} {\bfseries 311} no.~5764, (Feb, 2006) 1133--1135}. \url{https://doi.org/10.1126%2Fscience.1121541}.

\bibitem{https://doi.org/10.48550/arxiv.quant-ph/0502070}
M.~A. Nielsen, ``A geometric approach to quantum circuit lower bounds,'' 2005.
\newblock \url{https://arxiv.org/abs/quant-ph/0502070}.

\bibitem{https://doi.org/10.48550/arxiv.quant-ph/0701004}
M.~R. Dowling and M.~A. Nielsen, ``The geometry of quantum computation,'' 2007.
\newblock \url{https://arxiv.org/abs/quant-ph/0701004}.

\bibitem{Jefferson:2017sdb}
R.~Jefferson and R.~C. Myers, ``{Circuit complexity in quantum field theory},'' \href{http://dx.doi.org/10.1007/JHEP10(2017)107}{{\em JHEP} {\bfseries 10} (2017) 107}, \href{http://arxiv.org/abs/1707.08570}{{\ttfamily arXiv:1707.08570 [hep-th]}}.

\bibitem{Khan:2018rzm}
R.~Khan, C.~Krishnan, and S.~Sharma, ``{Circuit Complexity in Fermionic Field Theory},'' \href{http://dx.doi.org/10.1103/PhysRevD.98.126001}{{\em Phys. Rev. D} {\bfseries 98} no.~12, (2018) 126001}, \href{http://arxiv.org/abs/1801.07620}{{\ttfamily arXiv:1801.07620 [hep-th]}}.

\bibitem{Bhattacharyya:2018bbv}
A.~Bhattacharyya, A.~Shekar, and A.~Sinha, ``{Circuit complexity in interacting QFTs and RG flows},'' \href{http://dx.doi.org/10.1007/JHEP10(2018)140}{{\em JHEP} {\bfseries 10} (2018) 140}, \href{http://arxiv.org/abs/1808.03105}{{\ttfamily arXiv:1808.03105 [hep-th]}}.

\bibitem{Chapman:2018hou}
S.~Chapman, J.~Eisert, L.~Hackl, M.~P. Heller, R.~Jefferson, H.~Marrochio, and R.~C. Myers, ``{Complexity and entanglement for thermofield double states},'' \href{http://dx.doi.org/10.21468/SciPostPhys.6.3.034}{{\em SciPost Phys.} {\bfseries 6} no.~3, (2019) 034}, \href{http://arxiv.org/abs/1810.05151}{{\ttfamily arXiv:1810.05151 [hep-th]}}.

\bibitem{Caceres:2019pgf}
E.~Caceres, S.~Chapman, J.~D. Couch, J.~P. Hernandez, R.~C. Myers, and S.-M. Ruan, ``{Complexity of Mixed States in QFT and Holography},'' \href{http://dx.doi.org/10.1007/JHEP03(2020)012}{{\em JHEP} {\bfseries 03} (2020) 012}, \href{http://arxiv.org/abs/1909.10557}{{\ttfamily arXiv:1909.10557 [hep-th]}}.

\bibitem{Hackl:2018ptj}
L.~Hackl and R.~C. Myers, ``{Circuit complexity for free fermions},'' \href{http://dx.doi.org/10.1007/JHEP07(2018)139}{{\em JHEP} {\bfseries 07} (2018) 139}, \href{http://arxiv.org/abs/1803.10638}{{\ttfamily arXiv:1803.10638 [hep-th]}}.

\bibitem{Chapman:2017rqy}
S.~Chapman, M.~P. Heller, H.~Marrochio, and F.~Pastawski, ``{Toward a Definition of Complexity for Quantum Field Theory States},'' \href{http://dx.doi.org/10.1103/PhysRevLett.120.121602}{{\em Phys. Rev. Lett.} {\bfseries 120} no.~12, (2018) 121602}, \href{http://arxiv.org/abs/1707.08582}{{\ttfamily arXiv:1707.08582 [hep-th]}}.

\bibitem{Camargo:2018eof}
H.~A. Camargo, P.~Caputa, D.~Das, M.~P. Heller, and R.~Jefferson, ``{Complexity as a novel probe of quantum quenches: universal scalings and purifications},'' \href{http://dx.doi.org/10.1103/PhysRevLett.122.081601}{{\em Phys. Rev. Lett.} {\bfseries 122} no.~8, (2019) 081601}, \href{http://arxiv.org/abs/1807.07075}{{\ttfamily arXiv:1807.07075 [hep-th]}}.

\bibitem{Ali:2018aon}
T.~Ali, A.~Bhattacharyya, S.~Shajidul~Haque, E.~H. Kim, and N.~Moynihan, ``{Post-Quench Evolution of Complexity and Entanglement in a Topological System},'' \href{http://dx.doi.org/10.1016/j.physletb.2020.135919}{{\em Phys. Lett. B} {\bfseries 811} (2020) 135919}, \href{http://arxiv.org/abs/1811.05985}{{\ttfamily arXiv:1811.05985 [hep-th]}}.

\bibitem{Ali:2019zcj}
T.~Ali, A.~Bhattacharyya, S.~S. Haque, E.~H. Kim, N.~Moynihan, and J.~Murugan, ``{Chaos and Complexity in Quantum Mechanics},'' \href{http://dx.doi.org/10.1103/PhysRevD.101.026021}{{\em Phys. Rev. D} {\bfseries 101} no.~2, (2020) 026021}, \href{http://arxiv.org/abs/1905.13534}{{\ttfamily arXiv:1905.13534 [hep-th]}}.

\bibitem{Balasubramanian:2018hsu}
V.~Balasubramanian, M.~DeCross, A.~Kar, and O.~Parrikar, ``{Binding Complexity and Multiparty Entanglement},'' \href{http://dx.doi.org/10.1007/JHEP02(2019)069}{{\em JHEP} {\bfseries 02} (2019) 069}, \href{http://arxiv.org/abs/1811.04085}{{\ttfamily arXiv:1811.04085 [hep-th]}}.

\bibitem{Balasubramanian:2021mxo}
V.~Balasubramanian, M.~DeCross, A.~Kar, Y.~C. Li, and O.~Parrikar, ``{Complexity growth in integrable and chaotic models},'' \href{http://dx.doi.org/10.1007/JHEP07(2021)011}{{\em JHEP} {\bfseries 07} (2021) 011}, \href{http://arxiv.org/abs/2101.02209}{{\ttfamily arXiv:2101.02209 [hep-th]}}.

\bibitem{Chowdhury:2023iwg}
S.~Chowdhury, M.~Bojowald, and J.~Mielczarek, ``{Geometric quantum complexity of bosonic oscillator systems},'' \href{http://dx.doi.org/10.1007/JHEP10(2024)048}{{\em JHEP} {\bfseries 10} (2024) 048}, \href{http://arxiv.org/abs/2307.13736}{{\ttfamily arXiv:2307.13736 [quant-ph]}}.

\bibitem{Bhattacharyya:2024rzz}
A.~Bhattacharyya, S.~Brahma, S.~Chowdhury, and X.~Luo, ``{Benchmarking quantum chaos from geometric complexity},'' \href{http://dx.doi.org/10.1007/JHEP03(2025)177}{{\em JHEP} {\bfseries 03} (2025) 177}, \href{http://arxiv.org/abs/2410.18754}{{\ttfamily arXiv:2410.18754 [hep-th]}}.

\bibitem{AIF_1966__16_1_319_0}
V.~Arnold, ``Sur la g\'eom\'etrie diff\'erentielle des groupes de {Lie} de dimension infinie et ses applications \`a l'hydrodynamique des fluides parfaits,'' \href{http://dx.doi.org/10.5802/aif.233}{{\em Annales de l'Institut Fourier} {\bfseries 16} no.~1, (1966) 319--361}. \url{https://www.numdam.org/articles/10.5802/aif.233/}.

\bibitem{Balasubramanian:2019wgd}
V.~Balasubramanian, M.~Decross, A.~Kar, and O.~Parrikar, ``{Quantum Complexity of Time Evolution with Chaotic Hamiltonians},'' \href{http://dx.doi.org/10.1007/JHEP01(2020)134}{{\em JHEP} {\bfseries 01} (2020) 134}, \href{http://arxiv.org/abs/1905.05765}{{\ttfamily arXiv:1905.05765 [hep-th]}}.

\bibitem{Chowdhury:2024ntx}
S.~Chowdhury, M.~Bojowald, and J.~Mielczarek, ``{Geometric measure of quantum complexity in cosmological systems},'' \href{http://dx.doi.org/10.1103/PhysRevD.111.036036}{{\em Phys. Rev. D} {\bfseries 111} no.~3, (2025) 036036}, \href{http://arxiv.org/abs/2407.01677}{{\ttfamily arXiv:2407.01677 [quant-ph]}}.

\bibitem{Mufti}
A.~Mufti, H.~A. Schmitt, and I.~Sargent, M., ``Finite‐dimensional matrix representations as calculational tools in quantum optics,'' \href{http://dx.doi.org/10.1119/1.17149}{{\em American Journal of Physics} {\bfseries 61} no.~8, (08, 1993) 729--733}, \href{http://arxiv.org/abs/https://pubs.aip.org/aapt/ajp/article-pdf/61/8/729/12168870/729\_1\_online.pdf}{{\ttfamily https://pubs.aip.org/aapt/ajp/article-pdf/61/8/729/12168870/729\_1\_online.pdf}}. \url{https://doi.org/10.1119/1.17149}.

\bibitem{Gilmore1}
R.~Gilmore and J.~Yuan, ``Group theoretical approach to semiclassical dynamics: Multimode case,'' \href{http://dx.doi.org/10.1063/1.457142}{{\em The Journal of Chemical Physics} {\bfseries 91} no.~2, (07, 1989) 917--923}, \href{http://arxiv.org/abs/https://pubs.aip.org/aip/jcp/article-pdf/91/2/917/18981437/917\_1\_online.pdf}{{\ttfamily https://pubs.aip.org/aip/jcp/article-pdf/91/2/917/18981437/917\_1\_online.pdf}}. \url{https://doi.org/10.1063/1.457142}.

\bibitem{Ban1}
M.~Ban, ``Lie-algebra methods in quantum optics: The liouville-space formulation,'' \href{http://dx.doi.org/10.1103/PhysRevA.47.5093}{{\em Phys. Rev. A} {\bfseries 47} (Jun, 1993) 5093--5119}. \url{https://link.aps.org/doi/10.1103/PhysRevA.47.5093}.

\bibitem{Ban:93}
M.~Ban, ``Decomposition formulas for su(1, 1) and su(2) lie algebras and their applications in quantum optics,'' \href{http://dx.doi.org/10.1364/JOSAB.10.001347}{{\em J. Opt. Soc. Am. B} {\bfseries 10} no.~8, (Aug, 1993) 1347--1359}. \url{https://opg.optica.org/josab/abstract.cfm?URI=josab-10-8-1347}.

\bibitem{gilmore2006lie}
R.~Gilmore, {\em Lie Groups, Lie Algebras, and Some of Their Applications}.
\newblock Dover Books on Mathematics. Dover Publications, 2006.
\newblock \url{https://books.google.pl/books?id=N8UsAwAAQBAJ}.

\bibitem{Birrell:1982ix}
N.~D. Birrell and P.~C.~W. Davies, \href{http://dx.doi.org/10.1017/CBO9780511622632}{{\em {Quantum Fields in Curved Space}}}.
\newblock Cambridge Monographs on Mathematical Physics. Cambridge Univ. Press, Cambridge, UK, 2, 1984.

\bibitem{Schumaker:1986tlu}
B.~L. Schumaker, ``{Quantum mechanical pure states with gaussian wave functions},'' \href{http://dx.doi.org/10.1016/0370-1573(86)90179-1}{{\em Phys. Rept.} {\bfseries 135} no.~6, (1986) 317--408}.

\bibitem{Caves:1985zz}
C.~M. Caves and B.~L. Schumaker, ``{New formalism for two-photon quantum optics. 1. Quadrature phases and squeezed states},'' \href{http://dx.doi.org/10.1103/PhysRevA.31.3068}{{\em Phys. Rev. A} {\bfseries 31} (1985) 3068--3092}.

\bibitem{Albrecht:1992kf}
A.~Albrecht, P.~Ferreira, M.~Joyce, and T.~Prokopec, ``{Inflation and squeezed quantum states},'' \href{http://dx.doi.org/10.1103/PhysRevD.50.4807}{{\em Phys. Rev. D} {\bfseries 50} (1994) 4807--4820}, \href{http://arxiv.org/abs/astro-ph/9303001}{{\ttfamily arXiv:astro-ph/9303001}}.

\bibitem{Grishchuk:1990bj}
L.~P. Grishchuk and Y.~V. Sidorov, ``{Squeezed quantum states of relic gravitons and primordial density fluctuations},'' \href{http://dx.doi.org/10.1103/PhysRevD.42.3413}{{\em Phys. Rev. D} {\bfseries 42} (1990) 3413--3421}.

\bibitem{Grishchuk:1989ss}
L.~P. Grishchuk and Y.~V. Sidorov, ``{On the Quantum State of Relic Gravitons},'' \href{http://dx.doi.org/10.1088/0264-9381/6/9/002}{{\em Class. Quant. Grav.} {\bfseries 6} (1989) L161--L165}.

\bibitem{Grishchuk:1994sj}
L.~P. Grishchuk, ``{Density perturbations of quantum mechanical origin and anisotropy of the microwave background},'' \href{http://dx.doi.org/10.1103/PhysRevD.50.7154}{{\em Phys. Rev. D} {\bfseries 50} (1994) 7154--7172}, \href{http://arxiv.org/abs/gr-qc/9405059}{{\ttfamily arXiv:gr-qc/9405059}}.

\bibitem{Martin:2004um}
J.~Martin, ``{Inflationary cosmological perturbations of quantum-mechanical origin},'' \href{http://dx.doi.org/10.1007/11377306_7}{{\em Lect. Notes Phys.} {\bfseries 669} (2005) 199--244}, \href{http://arxiv.org/abs/hep-th/0406011}{{\ttfamily arXiv:hep-th/0406011}}.

\bibitem{Martin:2007bw}
J.~Martin, ``{Inflationary perturbations: The Cosmological Schwinger effect},'' \href{http://dx.doi.org/10.1007/978-3-540-74353-8_6}{{\em Lect. Notes Phys.} {\bfseries 738} (2008) 193--241}, \href{http://arxiv.org/abs/0704.3540}{{\ttfamily arXiv:0704.3540 [hep-th]}}.

\bibitem{Lemoine:2008zz}
M.~Lemoine, J.~Martin, and P.~Peter, eds., \href{http://dx.doi.org/10.1007/978-3-540-74353-8}{{\em {Inflationary cosmology}}}.
\newblock 2008.

\bibitem{Livine:2012mh}
E.~R. Livine and M.~Martin-Benito, ``{Group theoretical Quantization of Isotropic Loop Cosmology},'' \href{http://dx.doi.org/10.1103/PhysRevD.85.124052}{{\em Phys. Rev. D} {\bfseries 85} (2012) 124052}, \href{http://arxiv.org/abs/1204.0539}{{\ttfamily arXiv:1204.0539 [gr-qc]}}.

\bibitem{Mukhanov:2007zz}
V.~Mukhanov and S.~Winitzki, {\em {Introduction to quantum effects in gravity}}.
\newblock Cambridge University Press, 6, 2007.

\bibitem{Ford:2021syk}
L.~H. Ford, ``{Cosmological particle production: a review},'' \href{http://dx.doi.org/10.1088/1361-6633/ac1b23}{{\em Rept. Prog. Phys.} {\bfseries 84} no.~11, (2021) }, \href{http://arxiv.org/abs/2112.02444}{{\ttfamily arXiv:2112.02444 [gr-qc]}}.

\bibitem{Bhattacharyya:2020kgu}
A.~Bhattacharyya, S.~Das, S.~S. Haque, and B.~Underwood, ``{Rise of cosmological complexity: Saturation of growth and chaos},'' \href{http://dx.doi.org/10.1103/PhysRevResearch.2.033273}{{\em Phys. Rev. Res.} {\bfseries 2} no.~3, (2020) 033273}, \href{http://arxiv.org/abs/2005.10854}{{\ttfamily arXiv:2005.10854 [hep-th]}}.

\bibitem{Bhattacharyya:2020rpy}
A.~Bhattacharyya, S.~Das, S.~Shajidul~Haque, and B.~Underwood, ``{Cosmological Complexity},'' \href{http://dx.doi.org/10.1103/PhysRevD.101.106020}{{\em Phys. Rev. D} {\bfseries 101} no.~10, (2020) 106020}, \href{http://arxiv.org/abs/2001.08664}{{\ttfamily arXiv:2001.08664 [hep-th]}}.

\bibitem{Lehners:2020pem}
J.-L. Lehners and J.~Quintin, ``{Quantum Circuit Complexity of Primordial Perturbations},'' \href{http://dx.doi.org/10.1103/PhysRevD.103.063527}{{\em Phys. Rev. D} {\bfseries 103} no.~6, (2021) 063527}, \href{http://arxiv.org/abs/2012.04911}{{\ttfamily arXiv:2012.04911 [hep-th]}}.

\end{thebibliography}\endgroup
\bibliographystyle{utphys}

\end{document}